\documentclass[letterpaper,twocolumn,10pt]{article}
\usepackage{usenix2019}

\usepackage{graphicx}
\usepackage[font=small, labelfont=bf]{caption}
\usepackage{makecell}
\usepackage{pifont}
\usepackage{soul}
\usepackage{multirow,array}
\usepackage{tablefootnote}
\usepackage{array, boldline, booktabs}
\usepackage{txfonts}
\usepackage[inline]{enumitem}
\usepackage{xspace}
\usepackage{colortbl}
\usepackage{hhline}
\usepackage{breakurl}
\usepackage{lipsum,graphicx,subcaption}
\usepackage{float}

\usepackage{picins}
\usepackage{wrapfig}

\usepackage{url}

\usepackage{titlesec}
\titlespacing*{\section}{0pt}{0.5ex}{0.5ex}
\titlespacing*{\subsection}{0pt}{0.3ex}{0.3ex}
\titlespacing*{\subsubsection}{0pt}{0ex}{0ex}

\let\oldding\ding
\renewcommand{\ding}[2][1]{\scalebox{#1}{\oldding{#2}}}%

\newcommand\blfootnote[1]{%
  \begingroup
  \renewcommand\thefootnote{}\footnote{#1}%
  \addtocounter{footnote}{-1}%
  \endgroup
}

\usepackage{listings}
\usepackage{xcolor}
\usepackage{authblk}

\newcommand{\showComments}{yes}
\newcommand{\note}[2]{
    \ifthenelse{\equal{\showComments}{yes}}{\textcolor{#1}{#2}}{}
}

\newcommand{\rowan} {\textsc{Rowan}\xspace}
\newcommand{\rowankv} {\textsc{Rowan-KV}\xspace} 

\newcommand{\rpckv} {{RPC-KV}\xspace} 
\newcommand{\rwritekv} {{RWrite-KV}\xspace} 
\newcommand{\sharekv} {{Share-KV}\xspace} 
\newcommand{\batchkv} {{Batch-KV}\xspace} 

\newcommand{\kvput} {\texttt{PUT}\xspace} 
 
\newcommand{\kvget} {\texttt{GET}\xspace} 

\newcommand{\DLWA} {DLWA\xspace}

\newcommand{\smalltitle}[1]{
 \noindent
\textbf{#1.}
}

\definecolor{linecolor}{rgb}{0.38,0.157,0.404}
\newcommand{\codeline}[1]{\hyperref[fig:code]{\color{linecolor}{#1}}}

\newcommand{\rwrite} {\texttt{WRITE}\xspace} 
\newcommand{\rread} {\texttt{READ}\xspace} 
\newcommand{\rsend} {\texttt{SEND}\xspace} 
\newcommand{\rrecv} {\texttt{RECV}\xspace} 
\newcommand{\ratomic} {\texttt{ATOMIC}\xspace}
\newcommand{\rfaa} {\texttt{FETCH\_AND\_ADD}\xspace}

\begin{document}

\title{Replicating Persistent Memory Key-Value Stores\\ with Efficient RDMA Abstraction} 


\author[]{\rm Qing Wang}
\author[]{\rm Youyou Lu}
\author[]{\rm Jing Wang}
\author[]{\rm Jiwu Shu}

\affil[]{\textit{Tsinghua University}}


\maketitle

\begin{abstract}
    
    Combining persistent memory (PM) with RDMA is a promising approach to 
    performant replicated distributed key-value stores (KVSs).
    However, existing replication approaches do not work well when applied to PM KVSs:
    1) Using RPC induces software queueing and execution at backups,
    increasing request latency;
    2) Using one-sided RDMA \rwrite causes many streams of small PM writes, leading to severe device-level write amplification (\DLWA) on PM.

    In this paper, we propose \rowan, an efficient RDMA abstraction to handle replication writes in PM KVSs;
    it aggregates concurrent remote writes from different servers,
    and lands these writes to PM in a sequential (thus low \DLWA) and one-sided (thus low latency) manner.
    We realize \rowan with off-the-shelf RDMA NICs. 
    Further, we build \rowankv, a log-structured PM KVS using \rowan for replication.
    Evaluation shows that under write-intensive workloads, compared with PM KVSs using RPC and  RDMA \rwrite for replication, \rowankv boosts throughput by 1.22$\times$ and 1.39$\times$ as well as lowers median \kvput latency by 1.77$\times$ and 2.11$\times$, respectively,
    while largely eliminating \DLWA.

\end{abstract}

\section{Introduction}
\label{sec:intro}

\blfootnote{This is the pre-print version of our OSDI'23 paper, which has been accepted after experiencing the revise-and-submit process of OSDI'22.}
Replicated distributed key-value stores (KVSs) support many applications by providing durability and high availability~\cite{RAMCloudCase,SIGMOD21FoundationDB, SOSP07Dynamo}.
The recent commercialization of persistent memory (PM), e.g., Intel's Optane DIMMs,
enables local storage with extremely low latency (e.g., $\sim$100ns when persisting small data~\cite{FAST20Guide}).
When building replicated distributed KVSs with such fast storage media, 
network and CPU will become determinants of request latency,
since replicating an object (i.e., key-value pair) involves multiple times of network communication and request queueing/execution.

RDMA, a widely-deployed network technology~\cite{NSDI21Pangu, SICOMM16Roce, FacebookRDMA},  is promising to mitigate the network and CPU overhead.
First, RDMA delivers low latency ($\sim$2$\mu$s) due to protocol-offload RDMA NICs (RNICs) and  kernel-bypass software.
Second, RDMA provides one-sided \rwrite and \rread, allowing remote memory accesses without involvement of remote CPUs.
Recent work have leveraged \rwrite to replicate data in DRAM (i.e., \rwrite-enabled replication)~\cite{NSDI14FaRM,SOSP15FaRMTX,OSDI20Mu,OSDI18DrTMH}.
This eliminates software queueing/execution of backups in the critical path, thus significantly cutting the replication latency compared with RPC-enabled replication.

Yet, in the context of PM KVSs, \rwrite-enabled replication approach does not work well: it induces severe device-level write amplification (\DLWA) on PM.
Specifically, a KVS is typically finely sharded for load balancing and fast recovery, so every server acts as backups for many shards,
receiving numerous concurrent replication writes from many remote threads; besides, these replication writes are typically small ($\sim$100B) due to prevalent tiny objects in real-world workloads~\cite{FAST20FacebookRocksDB, SOSP21Kangaroo}.
In \rwrite-enabled replication approaches (e.g., FaRM~\cite{SOSP15FaRMTX}), each server allocates an exclusive backup log for \emph{every remote thread}, to accommodate remote \rwrite from primaries.
When adopting \rwrite-enabled replication to PM KVSs, 
these backup logs generate a huge number of PM write streams\footnote{A \emph{write stream} is a group of writes targeting contiguous addresses, e.g., writes that perform log appending.}, which contain lots of small-sized writes.
These numerous write streams lead to severe \DLWA, 
since PM has block access granularity at media level (e.g., 256B in Optane DIMMs) and 
its hardware combining capacity is bounded.
In our experiments, with 128B RDMA \rwrite, 144 remote PM write streams cause 1.58$\times$ \DLWA (\S\ref{moti:amplification}).
\DLWA wastes limited PM write bandwidth, shortens PM lifetime, and harms PM's persistence efficiency.

In this paper, we propose \rowan, an efficient RDMA abstraction to handle replication writes on PM KVSs.
\rowan can aggregate numerous concurrent remote writes from different servers, 
and land these writes to PM \emph{sequentially}, so as to largely eliminate \DLWA.
Besides, it is one-sided as RDMA \rwrite, enabling backup-passive replication with low latency and high CPU efficiency.
We realize \rowan with off-the-shelf RNICs based on two observations:
1) RDMA \rsend is two-sided on the control path but \emph{one-sided on the data path}; 
2) RNICs consume receive buffers \emph{in order}.
Thus, we let a control thread at the receiver side push PM-resident buffers into receive queues 
\emph{in increasing address order}.
Senders only need to issue \rsend for remote PM writes and wait for ACKs generated by receiver-side RNICs.
We leverage two RNIC hardware features, shared receive queue (SRQ)~\cite{SRQ} and multi-packet receive queue (MP RQ)~\cite{MPRQ,MPRQLinuxRDMA}, to merge writes from different connections and support variable-sized writes, respectively.
We also streamline \rowan's control path by minimizing the control thread's tasks.
A \rowan instance can achieve 54.5Mops/s for highly concurrent 64B remote PM writes, with almost no \DLWA.

Further, we build \rowankv, a PM KVS leveraging \rowan for primary-backup replication.
It adopts a log-structured approach to manage both local PM writes and remote PM writes.
Specifically, each server maintains per-thread primary logs and a \emph{single} backup log on PM.
For a \kvput request, a worker thread in servers generates a log entry containing the targeted object;
then, it persists the log entry into its local primary log via CPU instructions and every backup's backup log via one-sided \rowan.
For a \kvget request, the thread searches DRAM-resident indexes which point to objects in logs.
In this way, \rowankv features high performance and low \DLWA:
1) Replication bypasses CPUs of backups, ensuring low latency and saving CPU cycles for foreground operations;
2) The number of PM write streams in a server is small (i.e., $n$ primary logs + $1$ backup log, where $n$ is local thread count), enabling efficient write combining in PM hardware and thus largely eliminating \DLWA.
\rowankv also introduces a failover mechanism for fault tolerance and a dynamic resharding mechanism for load balancing.

We evaluate \rowankv on Optane DIMMs under a cluster of 14 machines (8 clients and 6 servers).
Our evaluation focuses on YCSB benchmarks~\cite{YCSB} with object sizes from three typical Facebook KVSs workloads~\cite{FAST20FacebookRocksDB} (i.e., ZippyDB, UP2X and UDB).
Compared with KVSs using RPC and \rwrite for replication,
\rowankv boosts throughput by 1.22$\times$ and 1.39$\times$, 
lowers median \kvput latency by 1.77$\times$ and 2.11$\times$, 
and lowers 99\% latency by 1.26$\times$ and 2.06$\times$, respectively, 
under write-intensive workloads.
In addition, the \DLWA is less than 1.032$\times$ in \rowankv, while 1.54$\times$ in the \rwrite-enabled KVSs.
Under read-intensive workloads, they have similar performance.
We also compare \rowankv with two software techniques mitigating \DLWA, i.e., batching and log sharing; \rowankv still outperforms them.

In summary, this paper makes the following contributions:

\begin{itemize}[ itemsep=0pt, parsep=0pt, labelsep=5pt, 
	leftmargin=*, topsep=0pt,partopsep=0pt]
\item It demonstrates that \rwrite-enabled replication can lead to severe device-level write amplification on PM KVSs.
\item It introduces \rowan abstraction and \rowankv with goals of low latency and low device-level write amplification.
\item It uses experiments to confirm the efficacy of \rowankv. 
\end{itemize}

\section{Background and Motivation}
\label{sec:moti}

In this section, we first provide the background on PM (\S\ref{moti:pm}) and RDMA (\S\ref{moti:rdma}). Then, we show that characteristics of typical KVSs architecture and workloads
together lead to high fan-in small writes for replication  (\S\ref{moti:fanin}).
Finally, with experiments, we demonstrate that when handling these writes, \rwrite-enabled replication causes severe \DLWA (\S\ref{moti:amplification}).

\subsection{Persistent Memory (PM)}
\label{moti:pm}

PM is a new kind of storage device that sits on the memory bus.
Thus, PM is byte-addressable and can be accessed by CPUs via \texttt{load}/\texttt{store} instructions.
In this paper, we focus on Intel's Optane DIMM, the only available PM product.

\smalltitle{PM performance}
Optane DIMMs have unique performance characteristics.
In terms of bandwidth, an Optane DIMM offers about 2GB/s for writes and 6GB/s for reads, which are 1/6 and 1/3 of DRAM, respectively.
In terms of latency, compared to DRAM, Optane DIMMs have similar write latency but 3$\times$ higher read latency~\cite{FAST20Guide}.
The limited write bandwidth and high read latency of Optane DIMMs are the main design  considerations for many PM systems~\cite{ASPLOS20FlatStore,OSDI20Assise, EuroSys21ChameleonDB, ATC21TIPS, OSDI21Nap,TOS21THDPMS,OSDI22ListDB}.

\begin{figure}[t!]

  \centering
  \includegraphics[width=0.85\linewidth]{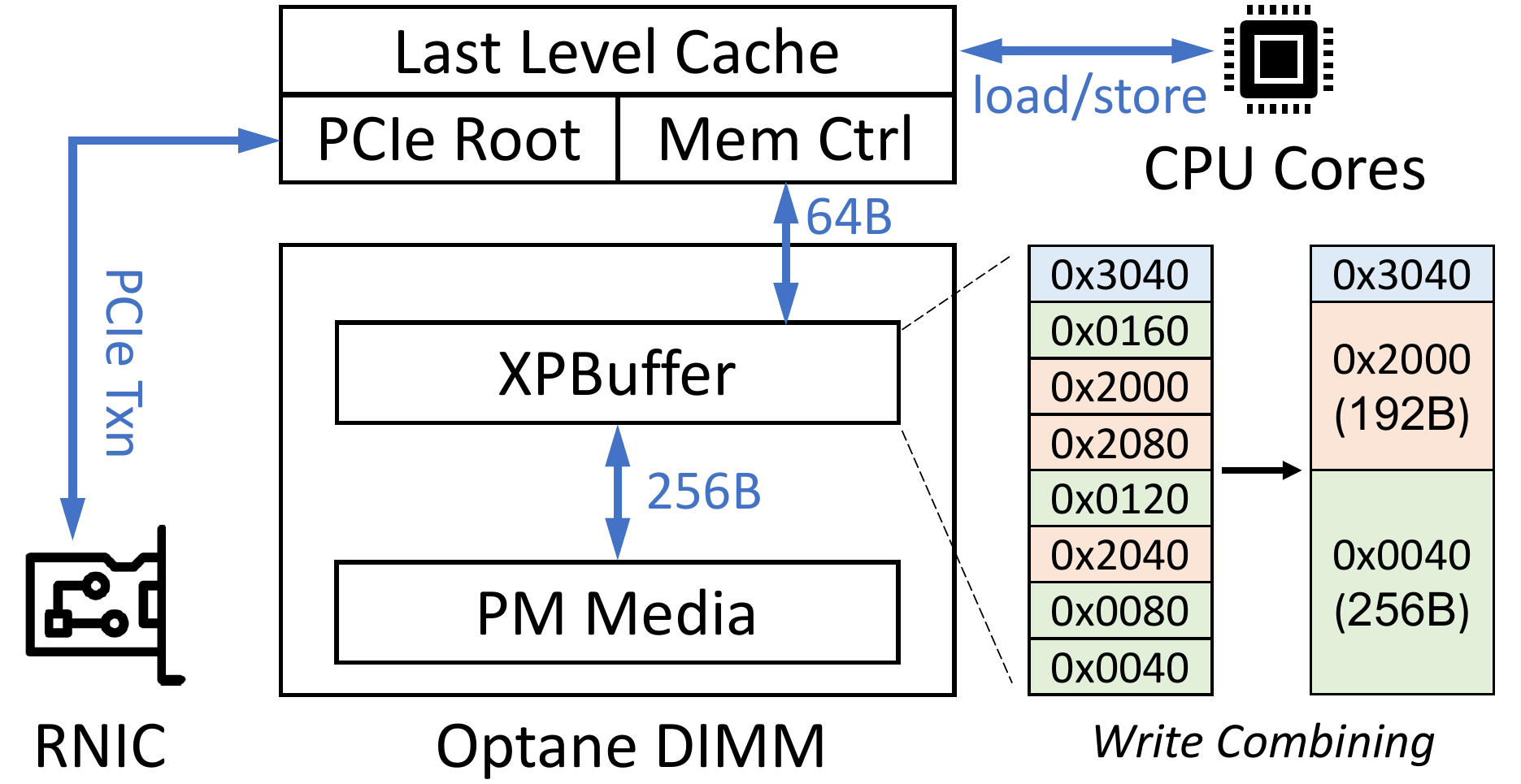}
  
  \vspace{-0.3cm}
  \caption{Architecture of Optane DIMMs and RNICs.}
  
  \vspace{-0.3cm}
  \label{fig:optane_arch}
\end{figure}

\smalltitle{PM architecture}
Figure~\ref{fig:optane_arch} presents the architecture of Optane DIMMs.
The memory controller generates cache-line granularity (i.e., 64B) read/write requests to Optane DIMMs, but the internal PM media has a 256B access granularity
(referred as \emph{XPLine} in this paper).
Such a granularity mismatch will trigger read-modify-write events, thus leading to 
device-level write amplification (\DLWA).
To mitigate \DLWA, each Optane DIMM features an XPBuffer~\cite{FAST20Guide}, which performs write combining for adjacent 64B writes, as shown in the right part of Figure~\ref{fig:optane_arch}.
Yang et al. estimated that the XPBuffer in an Optane DIMM is approximately 16KB in size~\cite{FAST20Guide}.

\smalltitle{Persistent modes}
There are two persistent modes for PM: ADR and eADR~\cite{VLDB20PMIdio}.
In ADR mode, once a \texttt{store} reaches the memory controller, it can survive power failure;
but the CPU cache is volatile, so programmers must explicitly flush data from the CPU cache (using \texttt{clwb} or \texttt{clflushopt} instructions) or bypass the CPU cache (using \texttt{ntstore} instructions).
In eADR mode, the CPU cache also belongs to the persistence domain: its data will be flushed to PM upon power failure.

\subsection{Remote Direct Memory Access (RDMA)}
\label{moti:rdma}

RDMA is a network technology that offers high bandwidth (e.g., 100 Gbps) and low latency
($\sim$2$\mu$s).

\smalltitle{Verb types}
RDMA provides two types of verbs for network communication: \emph{message verbs} and \emph{memory verbs}.
Message verbs, i.e., \rsend and \rrecv, are the same as Linux socket interfaces:
a \rsend emits a message to a remote server that prepares receive buffers via \rrecv.
Memory verbs include \rwrite, \rread and \ratomic. 
These verbs can operate receivers' memory without involving receivers' CPUs.
Due to the \emph{one-sided} feature, memory verbs enjoy low latency and high CPU efficiency.

\smalltitle{Queue pair}
RDMA servers use queue pairs (QPs) for communication.
A QP contains a \emph{send queue} (SQ) and a \emph{receive queue} (RQ).
A server posts requests, including \rsend, \rwrite, \rread, and \ratomic, to the send queue,
and posts \rrecv to the receive queue for accommodating incoming \rsend messages.
A send/receive queue is associated with a \emph{completion queue} (CQ), which generates completion signals for posted verbs.

\smalltitle{Remote persistence}
When issuing a \rwrite to remote PM, 
to ensure the data persistence,
we should take two extra actions.
\ding[1.2]{172} 
Since receiver-side RNICs return acknowledgements before data in \rwrite is DMA-ed to PM,
we should send a \rread (1B in arbitrary addresses)
to flush RNIC and PCIe buffers at the receiver side~\cite{SOCC20PMRDMA}.
These two verbs (i.e., \rwrite followed by \rread) can be posted in one request according to the ordering guarantee of RDMA~\cite{ATC21PMRDMA}.
\ding[1.2]{173}
We should disable \emph{Data Direct I/O (DDIO)}~\cite{DDIO,ATC20DDIO}, 
a technology of Intel CPUs that lets RNICs directly DMA data to last level cache (LLC).
In ADR mode, disabling DDIO  ensures that DMA-ed data can reach persistence domain.
In eADR mode, it avoids PM write amplification resulting from LLC's near-random eviction (64B cache line vs. 256B XPLine)~\cite{SOCC20PMRDMA, ATC21PMRDMA}.

\begin{figure*}[t!]

  \centering
  \includegraphics[width=0.85\linewidth]{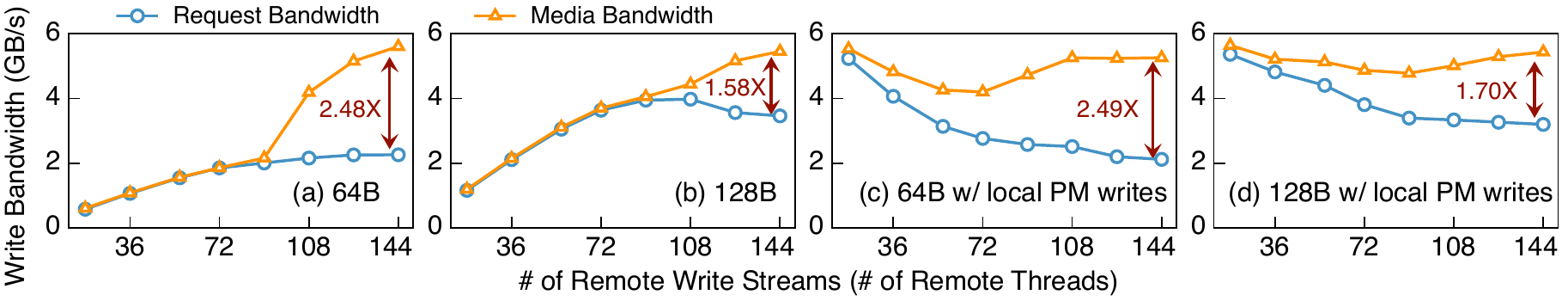}
  
  \vspace{-0.2cm}
  \caption{\DLWA with varying remote write streams. \emph{\DLWA$=media\ bandwidth/request\ bandwidth$. Threads access PM on a remote server; each thread generates a remote write stream. In (c) and (d), 18 CPU cores in the remote server perform local sequential PM writes.}}
  \vspace{-0.1cm}
  \label{fig:moti_wa}
\end{figure*}

\subsection{High Fan-in Small Writes in KVSs}
\label{moti:fanin}

\begin{table}[!t]
	\begin{center}		
	
		\resizebox{0.88\linewidth}{!}{
			\begin{tabular}{|l|c|c|}
              
        \cline{2-3}
				\multicolumn{1}{c|}{}
				 & \textbf{max shard size} & \makecell[c]{ \textbf{\# of backup shards}\\
                                        (stored by one PM server)} \\
				\hline
        \textbf{CosmosDB}& 20GB~\cite{CosmosDBShard}     & 200 \\
        
				\hline

        \textbf{DynamoDB}& 10GB~\cite{DynamoDBShard} & 400 \\
        \hline
        \textbf{FoundationDB}& 500MB~\cite{FoundationDBShard} & 8,400\\
        \hline
        \textbf{Cassandra}& 100MB~\cite{CassandraShard} & 42,000 \\
        \hline
        \textbf{TiKV}& 96MB~\cite{VLDB20TiDB} & 43,000\\
				\hline

			\end{tabular}
				}
    
		\caption{A PM server hosts many backup shards for popular KVSs.
    \emph{We assume 3-way replication and a typical configuration of PM servers: 2 sockets, each with 3TB Optane DIMMs (6TB in total). }}
    \vspace{-0.6cm}
    \label{tlb:fanin}
\end{center}
\end{table}

In KVSs,
replication makes \emph{high fan-in small writes} a dominant access pattern due to the following two reasons.

\smalltitle{1) Data sharding}
Distributed storage systems (including KVSs) typically split the entire data set into a large number of shards, and then distribute these shards across many servers~\cite{SOSP21ShardManager,OSDI16Slicer}.
Each shard has multiple replicas, with one selected as \emph{primary} and the others as \emph{backups}.
Data sharding has two advantages.
First, it can improve load balancing and support dynamic data migration in a fine-grained manner.
Second, it can improve availability:
when a server fails, since replicas of its data are distributed to many servers,
the system can perform recovery and re-replication in parallel.
For example, FaRM~\cite{NSDI14FaRM} maps each server into 100 consistent hashing rings by default;
in Facebook's RocksDB clusters, each server typically hosts tens or hundreds of shards~\cite{FAST21RocksDB}.





With data sharding, each server acts as backups for tens or hundreds of shards, and their primaries are distributed to many servers.
This makes every server receive messages for data replication, i.e., replication writes,
from many primaries residing in many other servers.
We call it \emph{high fan-in writes}.

To solidify the argument of high fan-in writes in KVSs, we analyze five widely-used replicated KVSs.
As shown in Table~\ref{tlb:fanin},  these KVSs all have a maximum shard size, from tens of megabytes (i.e., Cassandra~\cite{CassandraShard} and TiKV~\cite{VLDB20TiDB}) to several gigabytes (i.e., DynamoDB~\cite{DynamoDBShard} and CosmosDB~\cite{CosmosDBShard}).
When we deploy these KVSs on servers having terabytes of PM, 
each server will host a considerable number of backup shards 
which ranges from 200 (CosmosDB) to 43,000 (TiKV), 
generating high fan-in replication writes.

The degree of fan-in is even higher in systems equipped with fast network hardware (e.g., RNIC)~\cite{OSDI16FaSST, OSDI18DrTMH,EuroSys20Meerkat,ASPLOS20Hermes,NSDI14FaRM, SOSP15FaRMTX}.
To achieve multicore-scalable and squeeze out the raw performance of NICs, 
these systems run multiple threads, 
each independently processing requests using exclusive network connections.
For example, in DrTM+H~\cite{OSDI18DrTMH}, every worker thread independently issues RDMA \rwrite for replication.
With this threading model, the degree of fan-in increases from the number of remote servers to 
\emph{the number of remote threads}.

\smalltitle{2) Numerous small-sized objects}
Many important applications relying on KVSs generate numerous small objects,
whose size is much smaller than the access
granularity of PM media (e.g., 256B XPLine in Optane DIMMs).
For example, in ZippyDB, the largest KVS at Facebook~\cite{ZippyDBInfoQ}, the average size of objects is only 90.8B~\cite{FAST20FacebookRocksDB}.
Moreover, the other two typical KVSs at Facebook --- UP2X (a KVS for AI services) and  UDB (a KVS for social graph) --- have average object size of 57.25B and 153.8B, 
respectively~\cite{FAST20FacebookRocksDB}.
Twitter exhibits a similar workload feature: the most common length of a tweet is only 33 characters~\cite{TwitterSize, SOSP21Kangaroo}.
This paper focuses on these small objects because of their prevalence and importance.


When a KVS handles \texttt{PUT} requests 
(from clients) for these small objects,
primaries emit replication writes to associated backups.
These writes are small, since they typically only contain replicated objects with tiny metadata~\cite{RAMCloudCase}.
These writes are also high fan-in due to data sharding, as explained before.
As a result, we can conclude that \emph{high fan-in small writes are a
dominant access pattern in the cluster of KVSs}.



\subsection{\DLWA from \rwrite-enabled Replication} 
\label{moti:amplification}

Recent research demonstrates that for in-memory DRAM systems,
compared with RPCs,
leveraging RDMA \rwrite for replication can obtain significant performance gain~\cite{NSDI14FaRM, SOSP15FaRMTX, OSDI20Mu, OSDI18DrTMH}.
In such \rwrite-enabled replication, primaries issue replication writes to backups' logs via one-sided \rwrite, and only need to wait for acknowledgements (ACKs) from the RNIC hardware of backups.
This eliminates software queueing/execution of backups in the critical path, thus enjoying low latency (e.g., Mu~\cite{OSDI20Mu} cuts the latency by 61\%).
Further, the saved CPU cycles in backups can serve requests (e.g., \kvget) from clients, thus improving system throughput.

In systems using \rwrite-enabled replication,
to handle high fan-in replication writes from many remote threads (recall \S\ref{moti:fanin}),
each server maintains lots of backup logs, each accommodating \rwrite from
an individual remote thread (which can act as primary)~\cite{SOSP15FaRMTX,OSDI18DrTMH}.
For example, in FaRM's evaluation with 90 machines (each running 30 worker threads)~\cite{SOSP15FaRMTX},
there are thousands of backup logs (i.e., 89$\times$30) in each server.
Yet, when we apply \rwrite-enabled replication to PM KVSs,
these backup logs (which are placed in PM for durability) will cause a huge number of PM write streams,
which contain lots of small writes, thus inducing severe \DLWA.
We conduct an experiment to demonstrate it.



In the experiment, we launch a number of threads (on four servers), 
each issuing sequential RDMA \rwrite to an exclusive PM-resident log in a remote server and thus generating a PM write stream.
We disable DDIO in the remote server and post a \rread after each \rwrite to guarantee persistence.
The remote server is equipped with three 256GB Optane DIMMs and a 100Gbps RNIC.
We use \texttt{ipmctl}~\cite{ipmctl} to periodically read hardware counters of Optane DIMMs,
calculating \emph{request bandwidth} and \emph{media bandwidth}, which means write bandwidth  received from memory bus and write bandwidth issued to PM media, respectively.
Figure~\ref{fig:moti_wa}(a) and (b) show results with 64B and 128B \rwrite size 
(representing small replication writes, \S\ref{moti:fanin}), respectively.
When remote write stream count is lower than 90, \DLWA is negligible.
This is because the XPBuffer on Optane DIMMs can combine adjacent small writes from the same write streams into 256B internal writes (\S\ref{moti:pm}).
However, the capacity of combining is bounded due to the limited size of XPBuffer.
Consequently, as the number of remote write streams continues to increase, severe \DLWA appears.
Specifically, when remote write stream count is 144, the \DLWA is 2.48$\times$ and 1.58$\times$ in case of 64B \rwrite and 128B \rwrite, respectively.

Next, we consider a more practical scenario where local PM writes exist.
In the remote server, we run 18 CPU cores, each performing sequential 128B PM writes using \texttt{ntstore}.
We repeat the above experiment; Figure~\ref{fig:moti_wa}(c) and (d) show the results.
Without remote RDMA \rwrite, local PM writes can deliver high request bandwidth 
(i.e., available bandwidth).
As the remote write stream count increases, \DLWA in Optane DIMMs reaches 2.49$\times$ and 1.70$\times$ 
in case of 64B \rwrite and 128B \rwrite, respectively.
In addition, the available bandwidth drops 
from 5.2GB/s to 2.1GB/s (60\%) for 64B \rwrite, and 
from 5.4GB/s to 3.2GB/s (41\%) for 128B \rwrite. 

\DLWA on PM leads to three issues.
First, it reduces available PM write bandwidth, thus degrading system performance.
The wasted bandwidth could also have been used for 
co-located applications~\cite{ApSys21Dicio,OSDI20Caladan,NSDI19Shenango}.
Second, it shortens the lifetime of PM which has limited write endurance~\cite{OptaneEndurance}.
Third, severe \DLWA consumes a considerable number of hardware resources (e.g., XPBuffer), harming persistence efficiency.

To efficiently handle high fan-in small writes, we need a new RDMA abstraction (rather than \rwrite) for PM KVSs. This abstraction should \emph{mitigate \DLWA, while achieving benefits of one-sided verbs --- low latency and high CPU efficiency}.

\section{\rowan Abstraction}
\label{sec:rowan}

We propose \rowan,
a new RDMA abstraction to handle high fan-in small writes in PM KVSs.
In this section, we first describe \rowan's semantic and characteristics.
Then, we present how to realize \rowan using off-the-shelf RNICs. 

\begin{figure}[t!]

  \centering
  \includegraphics[width=0.6\linewidth]{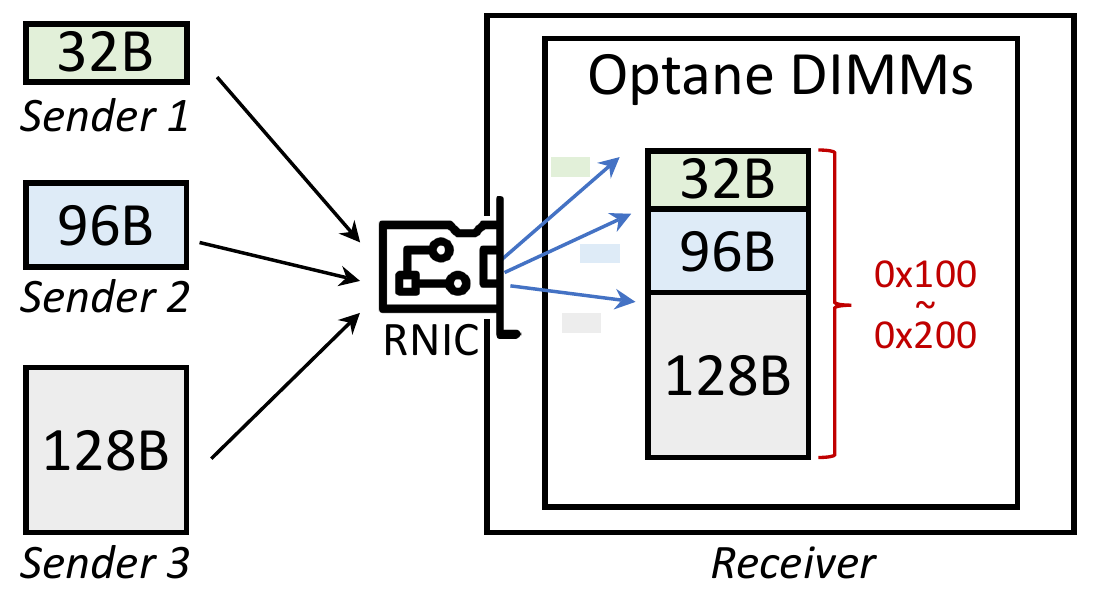}
  
  \vspace{-0.2cm}
  \caption{An instance of \rowan abstraction.}
  \vspace{-0.1cm}
  \label{fig:rowan_abstraction}
\end{figure}

\begin{figure*}[t!]

  \centering
  \includegraphics[width=0.85\linewidth]{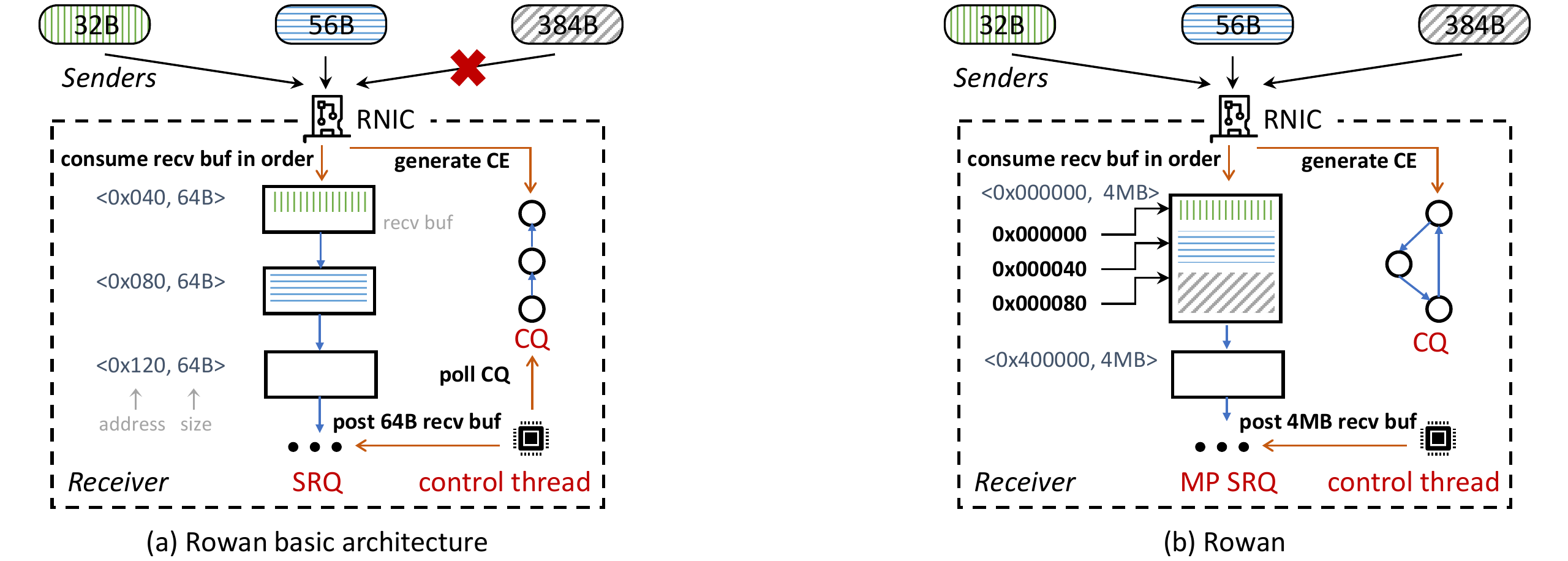}
  
  \vspace{-0.2cm}
  \caption{Realizing \rowan with off-the-shelf RNICs.
  \emph{\textbf{(a)} \rowan basic architecture using shared receive queue (SRQ). 
  In this subfigure, 
  the 32B write and 56B write are placed in the same XPLine. Yet, the 384B write fails to be received due to 64B receive buffers. 
  \textbf{(b)} \rowan using multi-packet shared receive queue (MP SRQ) with 64B stride. 
  In this subfigure, three writes are placed in two XPLines of the first receive buffer.
  We use a completion queue (CQ) ring to eliminate CQ polling in the control thread}.}
  \vspace{-0.2cm}
  \label{fig:rowanimpl}
\end{figure*}

\subsection{\rowan Semantic}

Figure~\ref{fig:rowan_abstraction} presents a \rowan instance.
A \rowan instance is associated with one receiver and a set of senders.
Senders concurrently issue writes to the receiver which has registered a large PM area. 
The receiver-side RNIC lands these writes to the PM area \emph{sequentially}, and finally returns ACKs to senders.

\rowan abstraction has the following advantages.
First, by translating concurrent remote small writes into a single write stream, 
the XPBuffer in Optane DIMMs can easily combine them into 256B XPLine writes,
largely eliminating \DLWA.
Second, since all the data operations are performed by the receiver-side RNIC without involving receiver-side CPUs,
\rowan enjoys benefits of low latency and high CPU efficiency like RDMA \rwrite.
In addition, compared with CPUs, RNIC ASICs can deliver extremely high throughput.

\smalltitle{Comparison with batching}
Batching is also an approach that can mitigate \DLWA on PM:
it opportunistically accumulates multiple small writes at the sender side, and then emits the batched writes to the receiver via one RDMA \rwrite.
However, batching induces extra latency, sapping the benefits of extremely low-latency hardware (i.e., RNICs and PM).
In contrast, \rowan does not delay any write and thus ensures low latency:
senders \emph{immediately} issue writes and 
receiver-side RNICs \emph{immediately} land received writes to PM.
In addition, as we will show in \S\ref{sec:eval}, 
batching frequently fails to accumulate enough small writes within a short time interval in KVSs,
and \rowan outperforms batching in both latency and throughput.
Our view of batching has been echoed by authors of RAMCloud --- 
\emph{``. . . batching requires some operations to be delayed until a full batch has been collected, and \textbf{this is not acceptable in a low-latency system} such as RAMCloud''}~\cite{RAMCloudCase}.



\subsection{High-Performance \rowan}
\label{subsec:rowan-impl}

\rowan is conceptually simple but challenging to realize using off-the-shelf RNICs.
We do not want to modify RNIC hardware like StRoM~\cite{EuroSys20StRoM} and PRISM~\cite{SOSP21PRISM}, so as to enable \rowan to be deployed immediately in datacenters today that are equipped with RNICs.
Before describing our solution, we present a straightforward solution that has poor performance.

\subsubsection{Straightforward Solution}
A straightforward solution to realize \rowan abstraction is combining RDMA \rwrite and atomic verb \rfaa.
Specifically, there is a 64-bit sequencer stored in the receiver's memory.
When performing a write, the sender first issues a \rfaa to the sequencer,
reserving a PM address; then, it issues a \rwrite to this address.
This solution has two limitations.
First, it needs two round trips, increasing the latency.
Second, the poor performance of atomic verbs bottlenecks throughput:
even storing the sequencer in RNICs' device memory~\cite{SIGMOD22Sherman}, the throughput is less than 10Mops/s.

\subsubsection{Our Solution}
Counter-intuitively, we use RDMA \rsend and \rrecv to realize \rowan.
This is based on our two observations.
\begin{itemize}[itemsep=0pt, parsep=0pt, labelsep=5pt, 
	leftmargin=*, topsep=0pt,partopsep=0pt]
\item \textbf{\emph{RDMA SEND is two-sided on the control path but one-sided on the data path}}.
In the control path, the receiver's CPUs prepare receive buffers via \rrecv;
however, in the data path, when handling \rsend requests, the receiver-side RNIC performs \emph{all} tasks,
including landing \rsend's data to receive buffers and returning ACKs.
\item \textbf{\emph{In a receive queue, receive buffers are consumed in order}}.
Every time, the receiver-side RNIC pops the \emph{first} buffer in the associated receive queue and
lands data to it.

\end{itemize}

\smalltitle{Key idea} On the control path, CPUs push PM buffers into the receive queue in increasing address order; on the data path,
the receiver-side RNIC consumes them in order.

\smalltitle{Basic architecture}
Figure~\ref{fig:rowanimpl}(a) shows the basic architecture of \rowan implementation.
\rowan uses reliable connection (RC) QPs to delegate transmission reliability to RNICs.
We create a shared receive queue (SRQ)~\cite{SRQ} which is associated with all QPs;
thus, RNICs can land data of \rsend from \emph{different remote QPs} to the same receive queue.
In the receiver, we reserve a dedicated thread, namely \emph{control thread},
to perform control-path tasks; the RNIC performs data-path tasks.

Specifically, the control thread splits the PM area into fixed-sized (e.g., 64B in Figure~\ref{fig:rowanimpl})(a)) buffers, and posts these buffers (using \rrecv) into the SRQ in \emph{increasing address order}.
Senders encapsulate writes into \rsend requests, and emit them to the receiver; each \rsend is followed by a \rread for persistence.
When receiving a \rsend (followed by a \rread), the receiver-side RNIC pops the first buffer in SRQ, DMAs the \rsend's data into the buffer, generates a completion entry (CE) to the SRQ's CQ, and finally returns an ACK to the sender. 
In this way, writes from different senders can be combined into the same XPLines on PM, mitigating \DLWA.

\smalltitle{Handling variable-sized writes}
When the size of a \rsend's data is larger than the first buffer in the SRQ,
the RNIC cannot accommodate it and will trigger an error CE.
For example, in Figure~\ref{fig:rowanimpl}(a), with 64B receive buffers,
the 384B write cannot be handled.
If we use a buffer size larger than 256B for the SRQ to support relatively large writes,
small writes from different senders will not be combined into the same XPLines, 
destroying the benefits of \rowan abstraction.

Fortunately, current RNICs (e.g., ConnectX-4/5/6) support a new type of RQ, called \emph{multi-packet receive queue}~\cite{MPRQ, MPRQLinuxRDMA} (MP RQ).
In an MP RQ, each receive buffer can accommodate \emph{multiple} \rsend requests.
We need to define a \emph{stride} (e.g., 64B) for an MP RQ.
When receiving a \rsend, the RNIC appends the data to the receive buffer that is being used, and the start address is stride-aligned.
If there is no enough space left, the RNIC pops a new receive buffer from the MP RQ to use.

Figure~\ref{fig:rowanimpl} shows \rowan that uses MP SRQ, 
where we set the stride to 64B and receive buffer size to 4MB.
In the figure, three writes are placed in two XPLines (i.e., 512B area) in the first receive buffer,
each having a 64B-aligned start address.
By using MP SRQ, \rowan can support variable-sized writes, while combining small writes to mitigate \DLWA.

There are two points worth noting when using MP SRQ:
\begin{itemize}[itemsep=0pt, parsep=0pt, labelsep=5pt, 
	leftmargin=*, topsep=0pt,partopsep=0pt]
\item In \rowan, the stride is a fixed value of 64B. 
We do not choose a smaller value (e.g., 32B) for two reasons.
First, in the RNIC we use (i.e., ConnectX-5), the minimum supported stride value is 64B. 
Second, recent studies suggest that senders should pad small writes to PCIe data word (64B) granularity~\cite{ATC21PMRDMA}, to avoid expensive read-modify-write operations on receivers' PM.
Thus, we assume the incoming small writes are already 64B granularity.

\item If a \rsend is larger than maximum transmission unit (MTU), it is comprised of multiple packets. The RNIC may land these packets to non-contiguous addresses.
We let the upper applications (e.g., KVSs) to handle this case.
\end{itemize}

\smalltitle{Minimizing control-path tasks}
On \rowan's data path, the receiver-side RNIC can deliver extremely high throughput (> 50Mops/s).
On the control path, for CPU efficiency, we only want to use \emph{one} control thread;
thus, we minimize control-path tasks to make them can be easily handled by one thread.

There are two tasks performed by the control thread: posting receive buffers into the MP SRQ and 
polling the CQ to consume CEs.
For the former, since we use large receive buffers (e.g., 4MB) by leveraging the multi-packet feature and post a batch of receive buffers at a time, this task is lightweight.
For the latter, unfortunately, unlike other verbs, \rrecv can not be marked as unsignaled,
so every \rsend will generate a CE at the receiver side.
The control thread cannot timely consume these CEs (considering > 50Mops/s throughput), making the CQ fill and thus causing QPs in an error state.
We get inspiration from eRPC~\cite{NSDI19eRPC} to address this problem.
Like eRPC, we create a CQ that forms a \emph{ring structure}, 
so that the RNIC can overwrite entries in the CQ ring in a round-robin manner.
In this way, the control thread does not need to poll the CQ.


\section{\rowankv Design}
\label{sec:kv}

We build \rowankv, a PM KVS that uses \rowan for primary-backup replication.
It has two main design goals.

\begin{itemize}[itemsep=0pt, parsep=0pt, labelsep=5pt, 
	leftmargin=*, topsep=0pt,partopsep=0pt]
\item \textbf{Low latency}. \rowankv exploits one-sided \rowan to eliminate software overhead at backups during replication.

\item \textbf{Low \DLWA}.
\rowankv adopts a log-structured approach to manage PM writes from both local CPUs and remote CPUs.
For the former, every thread appends data in its local log.
For the latter, \rowan merges replication writes into a \emph{single} backup log.
Hence, Optane DIMMs only receive a small number of write streams and can 
efficiently combine adjacent small writes into XPLines, thus mitigating \DLWA.

\end{itemize}

\begin{figure}[t!]

    \centering
    \includegraphics[width=\linewidth]{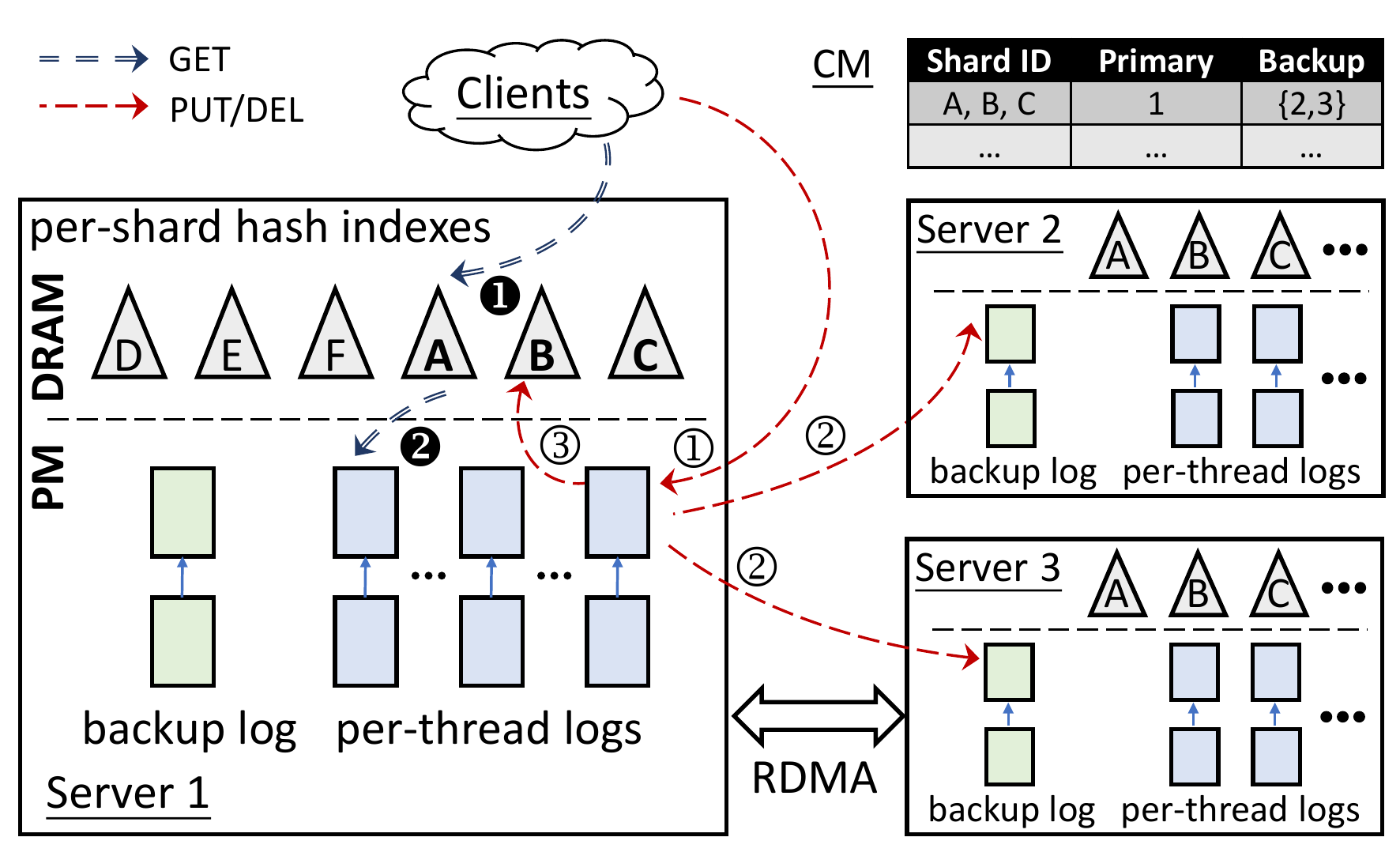}
    
    \vspace{-0.3cm}
    \caption{Architecture of \rowankv. \emph{The per-thread logs (t-logs) and backup log (b-log) are divided into 4MB segments.}} 
    \vspace{-0.2cm}
    \label{fig:rowankv}
  \end{figure}

  \subsection{Overview}

  Figure~\ref{fig:rowankv} shows the architecture of \rowankv.
  Servers persistently store objects (i.e., key-value pairs) in PM and use RDMA for network communication.
  \rowankv divides the entire data set into many shards and distributes them across servers. 
  Each shard is replicated for high availability: with the replication factor of \texttt{k}, it has one server as \emph{primary} and \texttt{k-1} servers as \emph{backups}.
  Clients issue KV requests via RPCs.

  \smalltitle{Sharding mechanism}
  \rowankv hashes each object's key into a 64-bit number and lets a shard manage a continuous  range in the hashed keyspace.
  Shard distribution is maintained by a \emph{configuration manager} (CM) and is cached in servers and clients.
  \rowankv uses a dynamic resharding mechanism to mitigate load imbalancing from overloaded servers (\S\ref{subsec:resharding}).

  \smalltitle{Log-structured approach}
  \rowankv adopts a log-structured approach, where each server has three components:

  \begin{itemize}[itemsep=0pt, parsep=0pt, labelsep=5pt, 
	leftmargin=*, topsep=0pt,partopsep=0pt]
\item \emph{Per-thread logs}.
Each server launches a number of \emph{worker threads} to handle requests from clients.
Each worker thread maintains a per-thread log (\textbf{t-log}) in PM, which stores objects of \texttt{PUT/DEL} requests.
We do not allocate independent logs for each shard, to reduce random PM writes.

\item \emph{Backup log}.
Each server has a \emph{single} backup log (\textbf{b-log}) in PM, which receives replication writes from primaries using a \rowan instance.
By doing so, \rowankv can largely eliminate \DLWA from high fan-in small writes. 

\item \emph{Per-shard hash indexes}.
Each server builds a DRAM-resident hash table for every shard it manages, 
to index objects in t-logs or the b-log.
Putting indexes in DRAM can avoid random PM writes and expensive PM reads~\cite{ASPLOS20FlatStore, VLDB21Viper}.

The t-logs and b-log are divided into 4MB \emph{segments}.

\end{itemize}

  \smalltitle{Handling KV requests}
  When issuing a KV request for an object, the client sends an RPC to a worker thread residing in the server that is the targeted shard's primary.

  For a \texttt{PUT/DEL} request, the worker thread generates a log entry containing the object (only the object's key for \texttt{DEL}),
  and persistently appends the log entry to its local t-log using \texttt{ntstore} instructions
  (\ding[1.2]{172} in Figure~\ref{fig:rowankv}).
  Then, the worker thread issues replication write for every backup via one-sided \rowan, persistently appending the log entry to every backup's b-log (\ding[1.2]{173}).
  Upon receiving all ACKs from backups' RNICs, the worker thread updates the associated index to make the object (in t-logs) visible
  (\ding[1.2]{174}), and finally returns a response to the client.
  \rowankv has a strong durability guarantee: when a client receives the response of a \texttt{PUT}/\texttt{DEL} request, its effects have been persisted on all replicas.

  For a \texttt{GET} request, the worker thread first locates the object by searching the associated index (\ding[1.2]{182}). Then, it copies the object's value from t-logs (\ding[1.2]{183})
  and replies to the client.

  \smalltitle{Background operations}
  \rowankv uses three types of threads to perform background operations.

  \begin{itemize}[itemsep=0pt, parsep=0pt, labelsep=5pt, 
	leftmargin=*, topsep=0pt,partopsep=0pt]
  \item \emph{Control thread}. One control thread performs control-path tasks for the \rowan instance (\S\ref{sec:rowan}). In \rowankv, it pushes free segments to the b-log via RDMA \rrecv, and hands used segments over to digest threads.
  \item \emph{Digest threads}. There are multiple digest threads. They digest used segments from the b-log. Specifically, they parse log entries and update associated indexes.
  \item \emph{Clean threads}. There are multiple clean threads. They garbage collect stale objects in segments (from worker threads or digest threads) to reclaim free PM space.
  \end{itemize}


\subsection{Log Metadata}

In \rowankv, t-logs and b-log are comprised of multiple segments, each storing a number of log entries.
We describe segment metadata and log entry metadata, respectively.

\subsubsection{Segment Metadata}
\label{subsub:segmeta}

A segment's metadata mainly includes its \emph{state}.
  At any given time, each segment is in one of four states:

  \begin{itemize}[itemsep=0pt, parsep=0pt, labelsep=5pt, 
	leftmargin=*, topsep=0pt,partopsep=0pt]
\item \emph{Free}. The segment can be allocated to t-logs by worker threads, the b-log by the control thread, or clean threads.
\item \emph{Using}. The segment is being used by t-logs, the b-log, or clean threads; it has space to store new log entries.

\item \emph{Used}. It has no space to store new log entries, and some of its log entries have \emph{not} been persisted on all replicas.

\item \emph{Committed}. It has no space to store new log entries, and all of its log entries have been persisted on all replicas.

\end{itemize}

In addition to the state, a segment has an extra metadata called \emph{owner}, 
indicating which type of thread allocates it (e.g., worker threads).
Each server maintains a PM array called \emph{segment meta table} to record metadata for all its segments.

\begin{figure}[t!]
  
  \centering
  \includegraphics[width=0.8\linewidth]{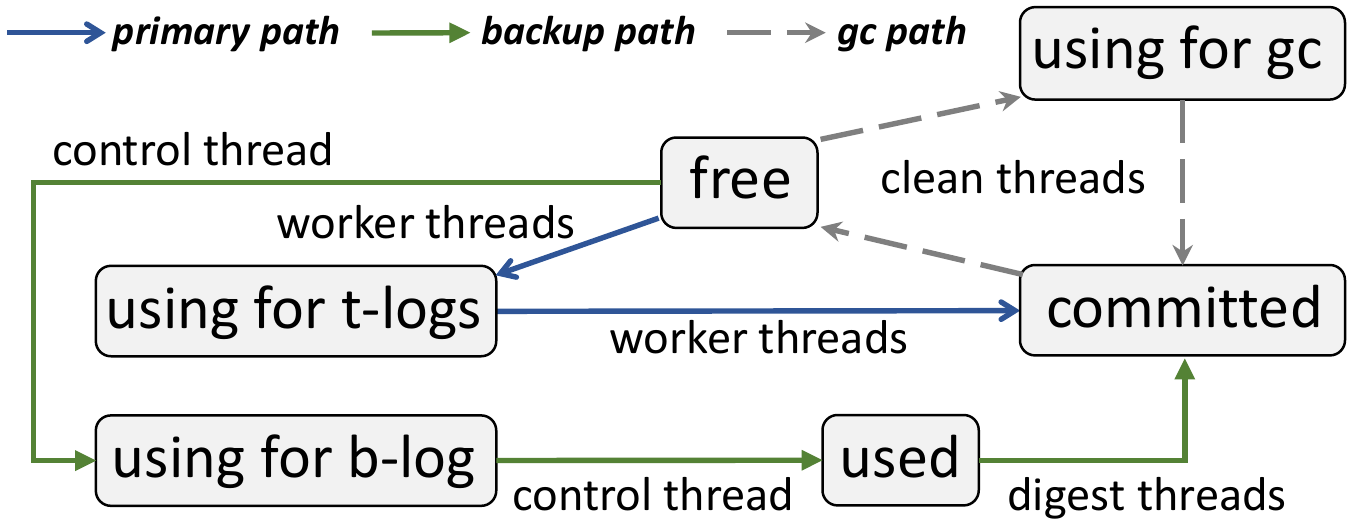}
  
  \vspace{-0.2cm}
  \caption{Life cycle of segments.}
  \vspace{-0.2cm}
  \label{fig:seglife}
\end{figure}

Figure~\ref{fig:seglife} presents the life cycle of segments.
The path for primaries is simple: 
a worker thread allocates a \emph{free} segment for its t-log, and the segment becomes \emph{using} state.
Once the segment has no space, it transitions into \emph{committed}, since the worker thread can easily ensure that all of the segment's log entries have been persisted on all replicas.
The path for backups is fairly complicated, 
where we should accurately distinguish between \emph{used} segments and \emph{committed} segments (\S\ref{subsec:manage-b-log} and \S\ref{subsec:digest}).
Such a distinguishment is essential for failover (\S\ref{subsec:failover}).

\subsubsection{Log Entry Metadata}

A log entry contains the request type (i.e., \texttt{PUT}/\texttt{DEL}) 
and the targeted object (only the object’s key for \texttt{DEL}).
It also includes three metadata fields:
\begin{itemize}[itemsep=0pt, parsep=0pt, labelsep=5pt, 
	leftmargin=*, topsep=0pt,partopsep=0pt]
    \item \emph{32-bit checksum}. The checksum covers the whole log entry.
    Checksums eliminate persistent tails for logs:
    upon recovery, we can identify the end of each log by calculating checksums.
    Besides, backups can use checksums to independently check the integrity of log entries in the b-log.

    \item \emph{48-bit version}. Each shard has a version, namely \emph{shard version},
    which is maintained by its primary.
    Upon a \texttt{PUT/DEL} request, the worker thread atomically increments the associated shard version, and stores the obtained version into the log entry.
    Upon recovery, the version allows us to identify the most recent objects from multiple t-logs.

    \item \emph{16-bit shard ID}. It indicates which shard the targeted object belongs to.
    
\end{itemize}

\begin{figure}[t!]
  \vspace{-0.3cm}
  \centering
  \includegraphics[width=0.9\linewidth]{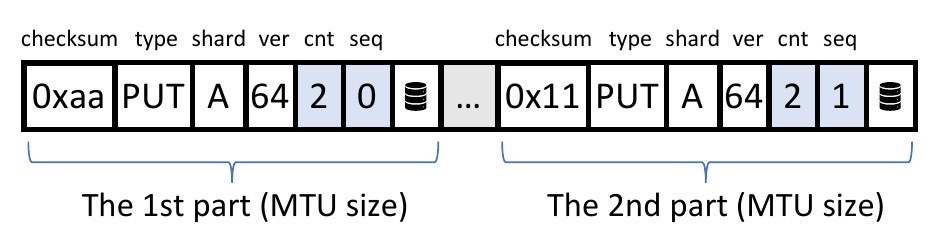}

  \vspace{-0.4cm}
  \caption{A 2-MTU-sized log entry in the b-log. 
  }
  \vspace{-0.2cm}

  \label{fig:MTU}
\end{figure}

\smalltitle{Handling larger-than-MTU log entries}
For a log entry that is larger than MTU, backup-side RNICs may divide it into multiple packets and place them in non-contiguous addresses of the b-log (recall \S\ref{subsec:rowan-impl}).
To enable backups check the integrity of such a log entry, we design a simple 
counter-based metadata.
Specifically, if a log entry is larger than MTU, 
we logically divide it into multiple MTU-sized blocks, and duplicate log entry metadata at the start of each block (each \emph{checksum} field protects the individual block).
Besides, we add two extra metadata to each block: 1) \emph{cnt}: block count of the log entry, 
and 2) \emph{seq}: the sequence number of the block.

Figure~\ref{fig:MTU} shows a 2-MTU-sized log entry in the b-log, where its two blocks are not adjacent. The pair of $\langle shard\ ID: A, version: 64\rangle$ uniquely identifies the log entry. When scanning the two blocks (checksums match) with their \emph{cnt} and \emph{seq}, 
backups can determine the log entry's integrity.

\subsection{Managing the Backup Log}
\label{subsec:manage-b-log}


The control thread manages the b-log by communicating with the RNIC and digest threads.
To minimize the communication overhead, the control thread performs tasks in a \emph{batch} manner.

Specifically, when the system starts up, the control thread allocates a considerable number of free segments (e.g., 512) for the b-log, and pushes them into \rowan's MP SRQ via \rrecv.
Then, it enters into a loop: 1) identifies a batch of segments (e.g., 128) that is in \emph{used} state; 2) hands these segments over to digest threads; 3) allocates a batch of free segments and pushes them into the b-log via one \rrecv call.
Note that a free segment transitions into the \emph{using} state after it is allocated by the control thread (recall backup path in Figure~\ref{fig:seglife}).

\smalltitle{Identifying used segments}
The control thread adopts a simple method to identify used segments in the b-log.
For every segment pushed into the b-log, its first 64 bits are set to zeros.
Meanwhile, the first 64 bits in a log entry include the request type, which is non-zero.
Thus, when the control thread finds that a segment has non-zero first 64 bits,
it can ensure that all \emph{previous} segments in the b-log (we call the set of segments \texttt{S} here) have been allocated by the RNIC for accommodating log entries.
However, this does not mean that segments in \texttt{S} are \emph{used}, 
since maybe some DMA operations writing log entries in \texttt{S} are outstanding.
Hence, we wait 2ms for all these DMA operations to complete, to guarantee that all segments in \texttt{S} have transitioned into \emph{used} state.
At the primary side, worker threads measure the time of each replication write:
if it is more than 1ms, worker threads retry the replication write.

\subsection{Digest and Garbage Collection}
\label{subsec:digest}

\smalltitle{Digest}
Multiple digest threads process used segments in the b-log in parallel.
Each digest thread manages an exclusive set of shards: 
it extracts log entries from used segments in order and only processes shards it manages. 
For a log entry, digest threads update the index of the associated shard.
Besides, digest threads identify \emph{committed} segments, and hand these segments over to clean threads.

\smalltitle{Identifying committed segments}
To help digest threads identify committed segments in the b-log, primaries disseminate the information of log entries to backups.
Specifically, for a shard, worker threads in its primary maintain a \texttt{\small CommitVer};
any log entry containing a version $\leq$ \texttt{\small CommitVer} has been persisted on all replicas.
Every 15ms, worker threads write the $\langle$shard ID, \texttt{\small CommitVer}$\rangle$ pair into backups' b-logs via \rowan.

At the backup side, digest threads maintain an array \texttt{\small CommitVerArray}, which contains 
associated \texttt{\small CommitVer} for each shard. 
When encountering a $\langle$shard ID, \texttt{\small CommitVer}$\rangle$ during parsing segments of the b-log, digest threads update \texttt{\small CommitVerArray}.
Meanwhile,
when processing a segment, digest threads generate an array \texttt{\small MaxVerArray} for it;
for each shard, this array records the \emph{maximum version} that digest threads have encountered  in log entries.
A used segment can transition into committed one, if its \texttt{\small MaxVerArray} $\leq$ \texttt{\small CommitVerArray} (i.e., for every shard, the maximum version in \texttt{\small MaxVerArray} $\leq$ \texttt{\small CommitVer} in \texttt{\small CommitVerArray}).

\smalltitle{Garbage collection}
Multiple clean threads garbage collect stale objects in committed segments.
When memory utilization of a committed segment, i.e., the percentage of valid bytes,
is lower than a pre-defined threshold (e.g., 75\% in our evaluation),
a clean thread cleans it.
Specifically, the clean thread scans the committed segment and checks the liveness of objects in log entries (by searching indexes).
For live objects, the clean thread copies associated log entries to a \emph{using} segment and updates indexes.
Finally, the committed segment transitions into \emph{free} state for future usages.

\subsection{Failover}
\label{subsec:failover}
We adopt FaRM's reconfiguration-style approach~\cite{SOSP15FaRMTX} to handle failover but tailor it for \rowankv.
A \emph{configuration} in \rowankv contains 1) 64-bit term, 2) membership, i.e., the set of live servers, and 3) shard distribution.
The configuration is persistently stored in a Zookeeper instance~\cite{ATC10Zookeeper},
and is cached in the CM, clients, and servers.
\rowankv uses leases to detect failures for servers and CM~\cite{SOSP15FaRMTX}.
When the CM fails, \rowankv activates a new CM using the same mechanism as FaRM~\cite{SOSP15FaRMTX}.
When a server fails, \rowankv performs failover with the following three phases.

\smalltitle{1) Generating and committing a new configuration}
The CM generates a new configuration, where the term is incremented and 
the membership excludes the failed server.
In the new shard distribution, the CM reassigns shards managed by the failed server to live servers,
and promotes a backup to the new primary for each shard losing its primary.

Then, the CM stores the new configuration in Zookeeper and sends it to all servers.
Servers cache the configuration, destroy QPs used for communicating with the failed server,
and respond.
From this point, servers block all requests from clients.
Once the CM receives all responses, after ensuring that the lease for the failed server has expired,
it sends a commit message to all servers.
Now, servers can unblock requests.
A server rejects requests containing terms that are lower than the one it caches.
Clients will fetch the new configuration from CM upon receiving rejected responses.

\smalltitle{2) Promoting backup to primary}
When a backup of a shard (we call the shard \texttt{A} here) is promoted to the new primary, 
its worker threads block requests to \texttt{A} until digest threads build indexes for all objects of \texttt{A}.
The new primary and backups should reach a consensus on the committed log entries.
Hence, the new primary and backups process \emph{using} and \emph{used} segments in the b-log, 
collecting log entries belonging to \texttt{A}.
These collected log entries are gathered to the new primary and then are scattered to backups.
The new primary and backups store these log entries into segments. In this way, all replicas 
will own the same set of log entries for \texttt{A}.
During digest, the new primary constructs a valid shard version for \texttt{A}, which is larger than versions in any \texttt{A}'s log entry.

\smalltitle{3) Re-replication}
The CM adds a new backup for the shard having replicas in the failed server.
The new backup performs re-replication asynchronously.
It first initializes an index for the shard, and then sends a message to the primary.
Upon receiving the message, the primary traverses the shard's index and transmits associated log entries to the new backup.

\subsection{Dynamic Resharding}
\label{subsec:resharding}
\rowankv introduces a dynamic resharding mechanism to migrate hotspot shards for improving load balancing. 

CM detects overloaded servers and produces new shard distribution.
Specifically,
for each shard, each worker thread records the number of received requests during a fixed period (i.e., 500ms), and sends the statistic data to CM.
Since \rowan is one-sided and thus backups are unaware of replication writes,
we let worker threads in primaries record the number of received replication writes for backup shards.
CM calculates the load of each server according to these statistics.
If a server has a load that is higher than the average load by a threshold (i.e., 30\%),
CM determines that the server is \emph{overloaded}.
CM produces a new shard distribution, where the hottest shards in overloaded servers are moved to underloaded servers, with a goal of making the load of every server within 5\% of the average.
Then, it saves a migration list in the configuration, which contains a triple
 $\langle$\emph{source server},  \emph{target server}, \emph{shard ID}$\rangle$ for each migration task.
Finally, CM increments the term, writes the new configuration (including the new shard distribution) to Zookeeper, and sends it to all servers.

Next, we describe how \rowankv migrates a primary shard from a \emph{source} server to a \emph{target} server (migrating a backup shard is much easier since it does not serve client requests).
Upon receiving the new configuration, servers cache it to local memory.
From this point, the source server rejects client requests for the migrated shard.
Clients will fetch the new configuration from CM when receiving rejected responses,
so subsequent requests to the migrated shard will be sent to the target server.
Then, the source server sends a message to the target server; the message contains the shard version of the migrated shard.
Upon receiving both the message and the new configuration, the targeted server starts to serve requests for the migrated shard.
In this way, \rowankv guarantees that only one server can serve the shard at any given time.
Then, the process of data migration starts:
\begin{itemize}[ itemsep=0pt, parsep=0pt, labelsep=5pt, 
	leftmargin=*, topsep=0pt,partopsep=0pt]
    \item In the source server, a migration thread requests free PM segments from the target server via RPCs, traverses the index of the migrated shard, and stores the associated log entries to remote segments via RDMA \rwrite.
  
    \item In the target server, a migration thread scans segments written by the source server and installs log entries in the shard's index.
    Upon a \kvput request to the migrated shard, the target server handles it as normal.
    Upon a \kvget request, the target server searches the index; 
    if the corresponding key is not found, the target server routes the \kvget request to the source server since some objects have not been migrated yet.
    Of note, the versions in log entries resolve the conflicts between the migration thread and concurrent \kvput requests.

\end{itemize}

The target server informs CM when it finishes data migration.
Then, CM deletes the migration task from the migration list and writes the new configuration to Zookeeper.
Finally, CM sends a message to the source server to inform it to free the index of migrated shard;
the associated log entries in the source server will be removed by garbage collection.

If the migration is interrupted due to failures of the source/target server, 
the CM first rolls back the shard distribution in the configuration to the state before migration.
Then, the CM deletes the associated task in the migration list and performs the normal failover process.
In addition, the CM informs the target server (if alive) to release resources allocated for the interrupted migration task (e.g., migration thread and index).

\subsection{Cold Start}
\label{subsec:coldstart}
When the entire cluster experiences a power failure,
\rowankv can guarantee durability of data.
Upon recovery, 
the CM fetches the configuration from Zookeeper, and disseminates it to all servers.
Each server obtains the metadata for all its segments via the segment meta table (recall \S\ref{subsub:segmeta}).
For a shard, its primary extracts associated log entries from \emph{using} segments whose \emph{owner} is worker threads;
then, the primary sends these log entries to backups, to make all replicas own the same set of log entries.
Each primary builds indexes for shards it manages by processing segments, 
and constructs valid shard versions.
If two log entries have the same targeted key, the one with the larger version is more recent.
Finally, \rowankv resumes unfinished migration tasks according to the migration list stored in the configuration. 

\section{Implementation}
\label{impl}

We implement \rowankv in Linux hosts.
\rowankv is a fully user-space system:
it uses \emph{libibverbs} for RDMA operations and 
CPU memory instructions for accessing PM.

\subsection{Threading Model}
\label{impl:threading}

\rowankv binds each thread (i.e., worker threads, clean threads, digest threads, and control thread) to an exclusive CPU core. \rowankv follows two principles:

\smalltitle{Minimizing inter-thread communication}
First, each worker thread handles both network I/O and KV logic;
this avoids request dispatch in systems that have dedicated threads to poll network requests~\cite{RAMCloudCase}, thus enjoying high multicore scalability.
Second, a thread hands over segments to other threads \emph{in a batch manner} (\S\ref{subsec:manage-b-log}) using thread-safe queues.

\smalltitle{Avoiding thread blocking}
To avoid blocking due to waiting for network events, 
worker threads adopt a coroutine-like approach to interleave
work:
after issuing \rowan operations for a \kvput, a worker thread saves the context of the \kvput request (e.g., the targeted key); 
then, it polls the RDMA completion queue, getting new requests to execute.
Upon receiving ACKs from backups, the worker thread restores the \kvput's context and continues the remaining logic.
In this way, a worker thread can concurrently handle multiple \kvput requests.

\subsection{Network Components}

\smalltitle{RPC}
\rowankv uses an RPC framework for client-server and inter-server communication (not include replication).
We build the RPC framework with RDMA \rsend and \rrecv verbs using unreliable datagram (UD) QPs.
Specifically, each worker thread creates a UD QP to receive requests and send responses.
When a client joins the \rowankv cluster, it establishes RPC connections with a worker thread in every server.
Like FaSST~\cite{OSDI16FaSST}, our RPC framework currently does not support messages larger than an MTU.
To reduce CPU consumption on PM reads:
the RPC framework leverages RNICs' scatter-gather DMA to gather RPC headers and PM-resident objects,
generating responses of \kvget requests.

\smalltitle{\rowan}
To realize \rowan,
every worker thread builds a reliable connection (RC) QP with every remote control thread.

At the sender side, a worker thread uses the associated send queue in QPs to issue \rowan operations.
A \rowan operation contains a \rsend followed by a 1B \rread for persistence (\S\ref{sec:rowan}).
\rsend and \rread are sent in one \texttt{ibv\_post\_send} call.
For a worker thread, all its \rowan QPs and RPC QP share the same CQ, 
so that it can be aware of \rowan ACKs and new RPC messages by polling the CQ.
We mark \rsend as unsignaled to eliminate a completion event.
For \rread, we store the context id of the associated \kvput request (\S\ref{impl:threading}) into the \texttt{wr\_id} field, so that worker threads can distinguish \rowan ACKs belonging to different \kvput requests when polling the CQ.

At the receiver side, a control thread manages all \rowan QPs connected to remote worker threads;
these QPs share an MP SRQ. The control thread pushes PM segments to the MP SRQ via \rrecv.
We register PM to RNICs using physical addresses~\cite{SOSP17LITE}, to remove virtual-to-physical translation tables in RNICs and thus reduce cache thrash of RNICs.

\smalltitle{Mitigating the impact of disabled DDIO}
We disable DDIO to ensure the RNICs can land data to PM (rather than CPU cache).
However, disabling DDIO will \ding[1.2]{172} cause CPU cache miss when handling RPCs and \ding[1.2]{173} degrade performance of DMA operations between RNICs and memory.
For \ding[1.2]{172}, worker threads poll multiple RPC messages at a time, and issue prefetch instructions to them.
For \ding[1.2]{173}, for RDMA \rread used for persistence, we set its source address to RNICs' device memory~\cite{DeviceMemory, SIGMOD22Sherman}, to eliminate a DMA write at senders.
We expect that DDIO does not need to be disabled, with next-generation RNICs supporting RDMA flush extensions~\cite{RDMAFlush}.

\subsection{Storage Components}

\smalltitle{PM management}
We configure Optane DIMMs in App-Direct mode, 
which exposes PM as a range of physical memory.
\rowankv splits the PM space into 4MB segments and stores the segment meta table in a predefined PM area (recall \S\ref{subsub:segmeta}).
A DRAM-resident free list records free segments, to serve segment allocation.
We add padding for each log entry, making it 64B-aligned;
it can \ding[1.2]{172} avoid expensive PM read-modify-writes on receiver-side RNICs~\cite{ATC21PMRDMA} when performing \rowan operations, and \ding[1.2]{173} avoid slow repeated writes to the same cache lines~\cite{ASPLOS20FlatStore, SOCC20PMRDMA} in logs.

\smalltitle{DRAM indexes}
Each per-shard index is implemented with a concurrent bucket hash table~\cite{NSDI14MICA}.
The hash table is organized into a bucket array, where each bucket contains multiple 64-bit items.
An item is composed of a 16-bit tag and a 48-bit PM address:
the tag is a part of a key's hash value, to filter out mismatched searches and thus reduce PM reads;
the PM address points to log entries.
For a key, its targeted bucket is calculated by {\small $hash(key)\ \%\ sizeof(bucket\ array)$}.
If the targeted bucket is full when inserting a key, threads create a new free bucket and link it to the targeted bucket, forming a bucket chain.
Indexes support conditional update to resolve conflicts between threads: indexes omit an update if its log entry has version that is smaller than the one indexes are pointing to.

\section{Evaluation}
\label{sec:eval}








\subsection{Experimental Setup}
\label{eval:setup}

\smalltitle{Environment}
We use 6 machines as servers and 8 machines as clients.
Each machine is equipped with the Intel Xeon Gold 6240M CPU (18 physical/36 logical cores),
96GB DRAM,
and one 100Gbps Mellanox ConnectX-5 RNIC.
All machines are connected to a 100Gbps Mellanox IB switch.
Each server machine owns three 256GB Optane DIMMs (ADR mode).

Unless otherwise specified, we run \rowankv on 6 servers. In each server, we use 24 cores for worker threads, 5 cores for digest threads, 6 cores for clean threads, and 1 core for control thread.
The control thread also manages leases, with a lease time of 10ms.
The CM and Zookeeper instance (3-way replication) run on client machines.
Each client machine runs multiple client threads to issue requests to servers.
We set the replication factor to 3.
Each server holds 48 shards.

\smalltitle{Workloads}
We evaluate \rowankv using YCSB~\cite{YCSB} with different \texttt{PUT}:\texttt{GET} ratios:
\textbf{Load A} --- 100\% \texttt{PUT} (write-only);
\textbf{A} --- 50\% \texttt{PUT} and 50\% \texttt{GET} (write-intensive);
\textbf{B} --- 5\% \texttt{PUT} and 95\% \texttt{GET} (read-intensive); 
\textbf{C} --- 100\% \texttt{GET} (read-only).
Key distribution follows Zipfian with parameter 0.99 (default parameter in YCSB).
We populate 200 million objects into KVSs before each experiment.
We use three Facebook workloads~\cite{FAST20FacebookRocksDB} to generate object size:
 \textbf{ZippyDB} (for general data store)  --- 90.8B average object size;
    \textbf{UP2X} (for  AI/ML services)  --- 57.25B average object size;
    \textbf{UDB} (for social graph)  --- 153.8B average object size.


  \begin{figure*}[!t]

    \centering
    \includegraphics[width=0.85\linewidth]{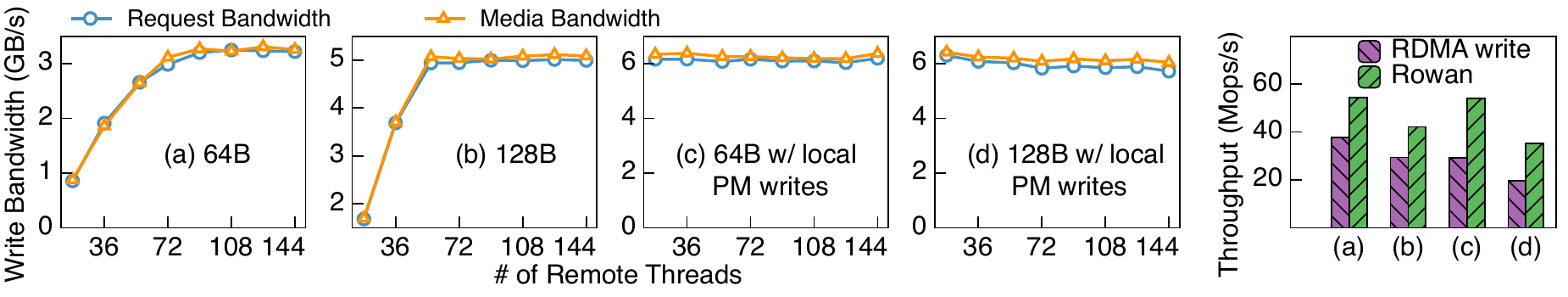}
    
    \vspace{-0.3cm}
    \caption{\rowan performance. \emph{\DLWA$=media\ bandwidth/request\ bandwidth$. A number of threads issue 64B/128B writes to a remote server's PM via a \rowan instance. In (c) and (d), 18 CPU cores in the remote server perform local sequential PM writes.}}
    \vspace{-0.3cm}
    \label{fig:rowan-eval}
  \end{figure*}

\smalltitle{Comparing targets}
We compare \rowankv with four KVSs, each using a specific replication approach:
\begin{itemize}[ itemsep=0pt, parsep=0pt, labelsep=5pt, 
	leftmargin=*, topsep=0pt,partopsep=0pt]
    \item \textbf{\rpckv}. It uses RPC to perform replication. Each server maintains per-thread b-logs, and primaries issue replication writes via RPC. Upon receiving a replication RPC, the worker thread appends the log entry into its local b-log.
    \item \textbf{\rwritekv}. It uses FaRM's approach~\cite{SOSP15FaRMTX} to perform replication. Each worker thread has an exclusive remote b-log at every remote server.
    During replication, the worker thread issues \rwrite for appending log entries to its b-logs at backups.
    Each server stores $(m-1) *n$ b-logs, where $m$ is the number of servers and $n$ is the worker thread count.

    \item \textbf{\batchkv}. \batchkv is a variant of \rwritekv and uses \rwrite for replication.
    Each worker thread generates large-sized \rwrite requests to its remote b-logs by \emph{batching log entries}, to mitigate \DLWA. To reduce latency, 
    worker threads immediately send batched log entries to backups once 1) the total size is larger than an XPLine, i.e., 256B, or 2) 5$\mu$s timeout is triggered.
    
    \item \textbf{\sharekv}. It is another variant of \rwritekv and uses \rwrite for replication. Worker threads in a server share the same remote b-log at a remote server, to reduce b-log count and thus mitigate \DLWA. 
    Worker threads use local atomic increment to obtain contiguous addresses in remote b-logs.

\end{itemize}

   All systems are implemented in the same codebase (including optimizations in \S\ref{impl}), to allow us to focus on the effects of replication approaches. By default, we disable DDIO to provide one-sided persistence. For \rpckv, DDIO is enabled.
   We compare \rowankv with two off-the-shelf KVSs in \S\ref{eval:existingKV}.

\subsection{\rowan Performance}
\label{eval:rowan}

We repeat the experiment in \S\ref{moti:amplification}, to show performance of \rowan abstraction.
Figure~\ref{fig:rowan-eval} presents the result of one \rowan instance.
\rowan can largely eliminate \DLWA in case of numerous concurrent remote small writes.
The \DLWA is less than 1.029$\times$ when no local PM writes exist (Figure~\ref{fig:rowan-eval}(a) and (b)), and less than 1.056$\times$ when local PM writes exist (Figure~\ref{fig:rowan-eval}(c) and (d))). This is because \rowan can merge remote small writes into a single write stream, enabling efficient hardware combining in Optane DIMMs' XPBuffer.

Further, we report the peak throughput of \rowan and RDMA \rwrite under these four cases,
as shown in the rightmost subfigure of Figure~\ref{fig:rowan-eval}.
When no local PM writes exist,
\rowan can deliver 54.5 Mops/s for 64B remote PM writes and 42.2 Mops/s for 128B one,
outperforming \rwrite by 1.44$\times$ and 1.43$\times$, respectively.
When local PM writes appear,
\rowan outperforms \rwrite by 1.85$\times$/1.78$\times$ for 64B/128B writes.
Three causes make \rowan performant.
First, \rowan largely eliminates \DLWA, 
improving the available PM bandwidth.
Second, on the data path of \rowan, all PM writes are performed by the receiver-side RNIC, 
ensuring high throughput.
Finally, on the control path, by leveraging ring CQ and MP SRQ, 
the control thread only performs very lightweight tasks, so it does not become the bottleneck.
Of note, 
the bottleneck of \rowan performance is 6GB/s PM write bandwidth in Figure~\ref{fig:rowan-eval}(b)-(c),
but processing capacity of RNICs in Figure~\ref{fig:rowan-eval}(a).
\rowan does not achieve 75Mops/s (a maximal message rate that a 100Gbps RNIC can provide),
since we disable DDIO and send an extra RDMA \rread for each \rowan operation.





\subsection{\rowankv Performance}
\label{eval:kv}

\begin{figure}[t!]

  \centering
  \includegraphics[width=0.9\linewidth]{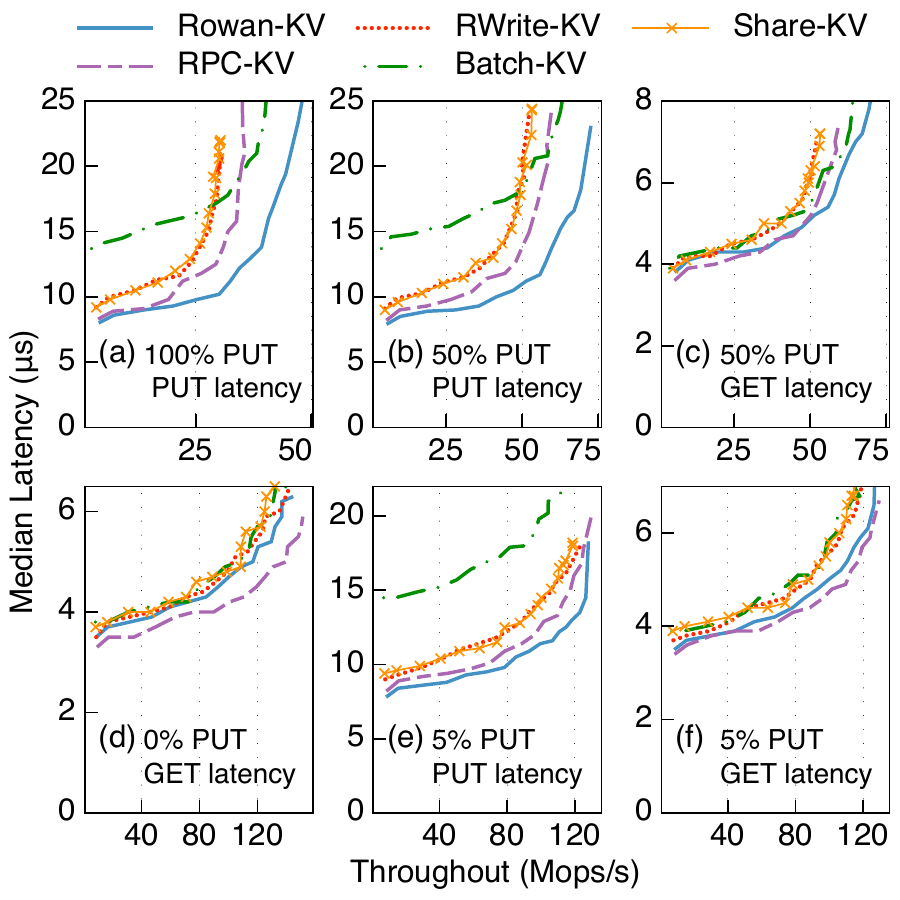}
   
  \vspace{-0.2cm}
  \caption{Median latency vs. throughput. \emph{ZippyDB object size.  We report \kvput latency and \kvget latency separately.}}
  \vspace{-0.2cm}
  \label{fig:ZippyDB-eval}
\end{figure}

Figure~\ref{fig:ZippyDB-eval} shows median latency and throughput (6 servers) under YCSB workloads with ZippyDB object size.
Since \rowankv aims to accelerate replication, we report latency of \kvput and \kvget separately.
We increase the load generated by clients, and ensure that KVSs reach their peak throughput.
We make two observations.

First, under read-only workloads (Figure~\ref{fig:ZippyDB-eval}(d)), \rpckv has 5\% higher throughput against other KVSs. This is because for KVSs using \rwrite or \rowan, DDIO is disabled, 
lowering RPC performance.
Such performance gap can be eliminated with next-generation RNICs supporting RDMA flush extensions~\cite{RDMAFlush}.
Under read-intensive workloads (Figure~\ref{fig:ZippyDB-eval}(e) and (f)), 
\rpckv and \rowankv have the similar throughput, 
since \rpckv consumes CPU cycles of backups for 5\% \kvput requests, offsetting the benefits of DDIO.
Compared with \rpckv, \rowankv has 1.09$\times$ lower median \kvput latency due to elimination of 
backups' software queueing, and 1.27$\times$ higher median \kvget latency due to disabled DDIO.

Second, under write-only and write-intensive (i.e., 50\% \kvput) workloads (Figure~\ref{fig:ZippyDB-eval}(a)-(c)), 
\rowankv has the highest throughput with the lowest median latency.
We compare \rowankv with the other four KVSs in turn.

\noindent
\underline{\emph{With \rpckv.}}  \rowankv achieves peak throughput of 72.7/48.2Mops/s under write-intensive/write-only workloads, outperforming \rpckv by 1.22$\times$/1.37$\times$. This is because \rowankv replicates log entries via \emph{one-sided} \rowan, saving CPU cycles that handle replication RPCs. 
The saved CPU cycles can be used for primaries to handle RPCs from clients.
At the peak throughput of \rpckv, \rowankv has 1.77$\times$/1.61$\times$ lower median \kvput latency under write-intensive/write-only workloads.
This is because compared with RPCs,  one-sided \rowan eliminates backup-side software 
queueing/execution on the critical path of replication.
Avoiding replication RPCs also makes \rowankv reduce median \kvget latency by 23\%.
Figure~\ref{fig:wa} shows \DLWA of write-only and write-intensive workloads 
(6 servers).
For \rowankv and \rpckv, the \DLWA is less than 1.032$\times$. 
This is because they generate a small number of PM write streams:
in each server, \rowankv has 24 t-logs and 1 b-log; \rpckv has 24 t-logs and 24 b-logs (recall we use 24 worker threads in experiments).
Optane DIMMs can efficiently combine adjacent small writes of the same logs into XPLine writes, when write stream count is not high (recall Figure~\ref{fig:moti_wa}(c) and (d)).

\begin{figure}[t!]

  \centering
  \includegraphics[width=0.85\linewidth]{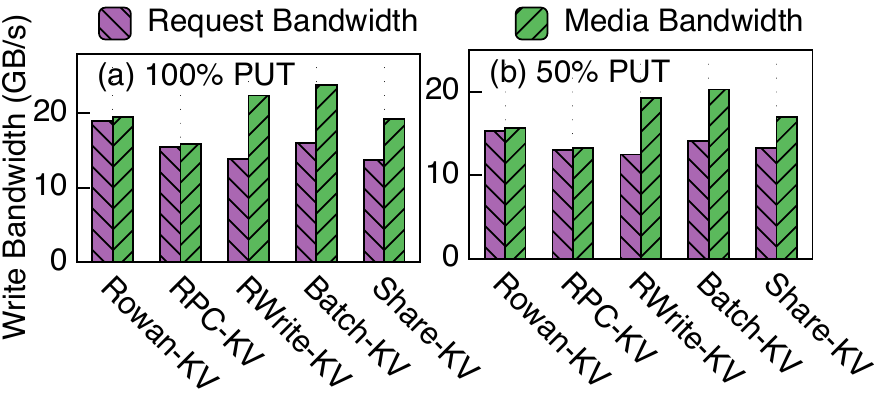}
  
  \vspace{-0.2cm}
  \caption{\DLWA (6 servers) at peak throughput.}
  \vspace{-0.2cm}
  \label{fig:wa}
\end{figure}

\underline{\emph{With \rwritekv.}} 
Compared to \rwritekv, \rowankv yields 1.39$\times$/1.61$\times$ higher throughput and
2.06$\times$/2.1$\times$ lower median \kvput latency under write-intensive/write-only workloads.
The main culprit of \rwritekv's low performance is \DLWA:
as shown in Figure~\ref{fig:wa}(a), it suffers 1.54$\times$ \DLWA.

\piccaption{Latency CDF. \label{fig:cdf}}\parpic[r]{\setcounter{figure}{10} \includegraphics[width=3.5cm]{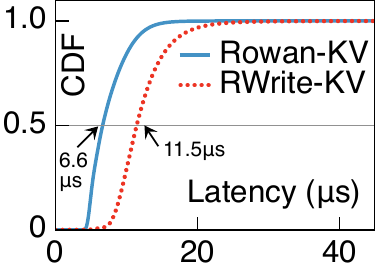} }
\noindent
This is because \rwritekv owns lots of logs (i.e., 24$\times$6 in experiments) in a server to accommodate small writes,  
exceeding the combining capacity of Optane DIMMs:
a large number of write streams are equivalent to random writes.
In \rwritekv, Optane DIMMs trigger lots of read-modify-write events, which squander a considerable number of hardware resources (e.g., XPBuffer), degrading performance of PM accesses.
To demonstrate it, we measure the latency of remote persistence operations of \rowankv and \rwritekv under write-intensive workloads.
Figure~\ref{fig:cdf} shows the latency distribution.
Remote persistence in \rwritekv is slow (against \rowankv), with 11.5$\mu$s median latency and 24$\mu$s 99\% latency.
Of note, 
although RNICs are ideally capable of providing an RTT of $\sim$2$\mu$s,
the 6.6$\mu$s median latency of \rowan is reasonable, since 1) 
we disable DDIO and each \rowan operation contains a synchronous RDMA \rread,
and 2) the latency is measured under high loads where RNICs suffer from DMA queueing.

\noindent
\underline{\emph{With \batchkv.}}  \batchkv boosts the throughput of \rwritekv by 1.23$\times$/1.35$\times$ under write-intensive/write-only workloads, since it
reduces the number of \rwrite and mitigates \DLWA (by 12\%) via batching. 
However, batching makes \batchkv suffer the highest \kvput latency among all KVSs:
even under low loads, \batchkv has more than 50\% higher \kvput latency compared with \rowankv.
In terms of throughput, \rowankv outperforms \batchkv by 1.13$\times$/1.19$\times$ under write-intensive/write-only workloads.
This is because \batchkv still experiences \DLWA:
it frequently fails to accumulate enough small writes within
5$\mu$s timeout for two reasons:
1) All \kvget requests do not generate writes but consume CPU time;
2) Only writes to the same destination can be batched; yet, due to sharding of KVSs, for a server acting as primaries, the backups of its shards are distributed to \emph{multiple} servers, greatly decreasing the batching opportunity.
We also change the timeout value to 20$\mu$s, and \batchkv delivers 9\% lower throughput against \rowankv, with unacceptable latency.



  

\noindent
\underline{\emph{With \sharekv.}} 
\sharekv reduces \DLWA of \rwritekv by 26\%/22\% under write-intensive/write-only workloads, since it lets worker threads share the same b-logs.
However, it still suffers sizable \DLWA (1.28$\times$$\sim$1.39$\times$), resulting in lower performance against \rowankv.
This is because 
although worker threads in a \sharekv server generate contiguous remote addresses for \rwrite, the asynchronous network makes receiver-side RNICs receive and perform these writes in an out-of-order manner.
In contrast, for \rowankv, leveraging \rowan, receiver-side RNICs decide destination addresses of writes.
Besides, \rowan can merge writes from \emph{different servers}.


\smalltitle{Tail latency}
Under write-intensive workloads with 50Mops/s throughput, \rowankv's 99\% latency is 20.5$\mu$s,
which is 1.26$\times$, 2.11$\times$, 1.53$\times$, and 1.87$\times$ lower than that of 
\rpckv, \rwritekv, \batchkv, and \sharekv, respectively.

\smalltitle{Performance under uniform workloads}
We evaluate \rowankv using uniform key distribution.
\rowankv delivers 67.86Mops/s and 108.19Mops/s in cases of 50\% \kvput and 5\% \kvput, respectively, which are 6.6\% and 15.5\% slower than throughput of Zipfian skewed workloads (see Figure~\ref{fig:ZippyDB-eval}).
\rowankv has higher performance under skewed workloads for two reasons.
First, in our cluster of 6 servers, due to hash-based sharding, 
there is no observable load imbalancing across servers under skewed workloads.
Second, threads enjoy much better cache locality under skewed workloads.


\begin{table}[!t]
	\begin{center}		
	
		\resizebox{\linewidth}{!}{
			\begin{tabular}{|c|c|c|c|c|c|}
              
        \cline{2-6}
				\multicolumn{1}{c|}{}
				 & \rowankv  & \rpckv & \rwritekv & \batchkv & \sharekv \\
				\hline
        \textbf{UP2X}& 73.9{\footnotesize Mops/s}        & 61.5{\footnotesize Mops/s}  & 56.2{\footnotesize Mops/s}  & 70.3{\footnotesize Mops/s}  & 56.0{\footnotesize Mops/s}  \\
        
				\hline

        \textbf{UDB}& 62.5{\footnotesize Mops/s} & 50.4{\footnotesize Mops/s} & 49.9{\footnotesize Mops/s} & 57.1{\footnotesize Mops/s} & 50.6{\footnotesize Mops/s} \\
        
				\hline

			\end{tabular}
				}
    
    \vspace{-0.2cm}
		\caption{Throughput under write-intensive workloads.
    \vspace{-0.4cm}
    \label{tlb:tp}
		}
\label{table:comparison}

\end{center}
\end{table}
\smalltitle{Performance with UP2X/UDB object size}
Due to space limitations, here we only report the throughput under write-intensive workloads,
as shown in Table~\ref{tlb:tp}.
\rowankv delivers the highest throughput via powerful \rowan abstraction.

\begin{figure*}[!h]

  \centering
  \includegraphics[width=0.83\linewidth]{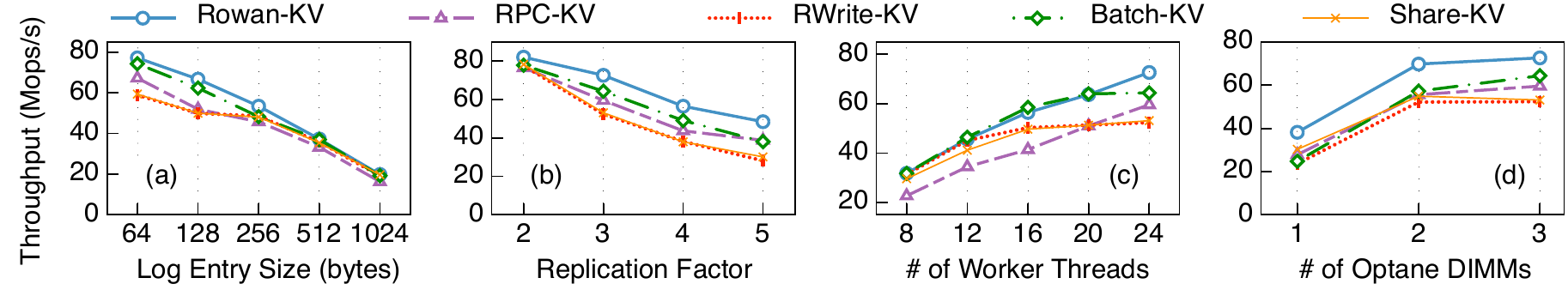}
  \setcounter{figure}{12} 
  \vspace{-0.3cm}
  \caption{Sensitivity analysis. \emph{We use write-intensive workloads with ZippyDB object size.}}
  \vspace{-0.4cm}
  \label{fig:sen}
\end{figure*}

\subsection{Sensitivity Analysis}
\label{eval:sens}

We conduct experiments on sensitivity analysis using write-intensive workloads 
and ZippyDB object size.

\smalltitle{Impact of object size}
We change object size to generate varying log entry size.
As shown in Figure~\ref{fig:sen}(a),
when log entry size is an integer multiple of XPLine size (e.g., 256B), 
all KVSs do not induce severe \DLWA;
thus, \rwritekv and KVSs using \rwrite have the similar throughput.
\rpckv consumes CPU cycles for replication RPCs, so it has 21\% lower throughput against \rowankv with 1024B log entries.

\smalltitle{Impact of replication factor}
Figure~\ref{fig:sen}(b) presents throughput with varying replication factor.
As replication factor increases,
performance improvement between \rowankv and other KVSs increases.
This is because, with higher replication factors, \rpckv needs to consume more CPU cycles to handle a \kvput request, and \rwrite-enabled KVSs issue more \rwrite and thus induce more \DLWA.
In contrast, \rowankv replicates objects in a one-sided manner and merges all remote writes into a single b-log in a sequential manner.

\begin{figure}[t!]
  \setcounter{figure}{13}
  \centering
  \includegraphics[width=0.83\linewidth]{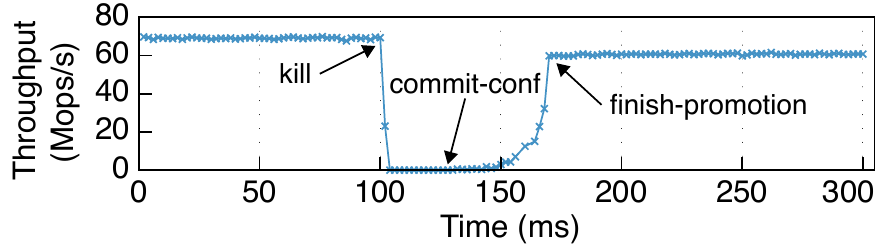}
  
  \vspace{-0.4cm}
  \caption{The timeline of failover.}
  \vspace{-0.3cm}
  \label{fig:failover}
\end{figure}

\smalltitle{Impact of worker thread count}
Figure~\ref{fig:sen}(c) presents
throughput with different worker thread counts.
We make two observations.
First, when the number of threads is small (i.e., $\le$ 16),
\rpckv has the lowest throughput, 
since the CPU is the bottleneck.
Second, \rwritekv and its variants yield poor scalability.
This is because 1) for \rwritekv and \batchkv, 
the number of b-logs is proportional to thread count,
and 2) for \sharekv, RNICs are more likely to receive and perform \rwrite to b-logs in an out-of-order manner in case of high thread count;
thus, they suffer more severe \DLWA with higher thread count.
\DLWA harms PM performance (recall Figure~\ref{fig:cdf}), thus stalling throughput.
In contrast, \rowankv exhibits superior throughput with different thread counts.

\smalltitle{Impact of PM bandwidth}
Figure~\ref{fig:sen}(c) presents
throughput with different number of Optane DIMMs per server.
In case of one Optane DIMM, the PM bandwidth is bottleneck.
Thus, \rwritekv (which has the most severe \DLWA) is outperformed by \rowankv, \rpckv, \batchkv, and \sharekv by 1.61$\times$, 1.18$\times$, 1.05$\times$, and 1.28$\times$, respectively.
In case of three Optane DIMMs, CPU becomes the bottleneck and limits throughput,
and PM bandwidth is not saturated (see Figure~\ref{fig:wa}).
\rowankv squeezes out CPU resources in two aspects: 1) it reduces CPU involvement via \rowan's one-sided semantic; 2) it largely eliminates \DLWA, 
streamlining Optane DIMMs' internal operations and thus improving persistence efficiency of worker threads.

\subsection{Failover and Cold Start}
\label{eval:failover}

\smalltitle{Failover}
We kill a server to test \rowankv's failover mechanism.
We use write-intensive workloads with ZippyDB objects and \rowankv runs for 50 seconds before the test.
Figure~\ref{fig:failover} shows the timeline, where throughput is recorded per 2ms.
The server is killed at time 100ms (i.e., \texttt{kill}).
\rowankv uses 26ms to commit the new configuration (i.e., \texttt{commit-config}), 
which mainly includes detecting failure (8ms), writing new configuration to Zookeeper (4.3ms),
and waiting for the failed server's lease to expire (10ms).
Then, \rowankv consumes about 44ms to promote backups to primaries 
(i.e., \texttt{finish-promotion}).
At this point, \rowankv can serve all requests from clients.

\smalltitle{Cold start}
We test cold start of a \rowankv instance, which contains 10 billion ZippyDB objects and thus occupies about 3TB PM space (6 servers). The time of cold start is 49.6s. 
Although cold start is slow, it is not common in datacenters.
Periodically checkpointing DRAM-resident indexes can accelerate cold start, and we leave it for future work.

\begin{figure}[t!]
  \centering
  \includegraphics[width=0.83\linewidth]{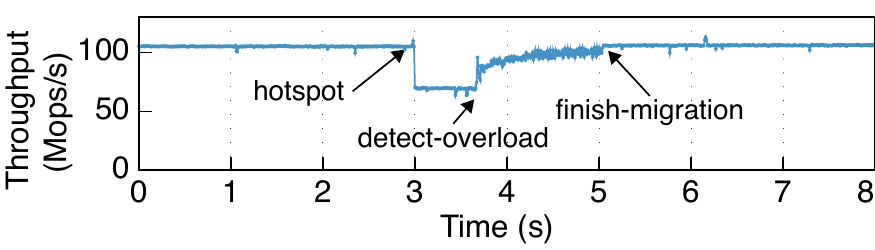}
  
  \vspace{-0.4cm}
  \caption{The timeline of resharding.}
  \vspace{-0.4cm}
  \label{fig:resharding}
\end{figure}

\subsection{Dynamic Resharding}
\label{eval:reshard}

In this experiment, we evaluate \rowankv's dynamic resharding mechanism.
We use read-intensive workloads with ZippyDB objects.
Figure~\ref{fig:resharding} presents the total throughput (6 servers) over time.
At first, clients generate a uniform workload and each server has a similar CPU utilization (i.e., 90.2\% $\sim$ 90.9\%).
At time 3s (i.e., \texttt{hotspot}), clients shift 80\% requests for server \texttt{A} to a shard residing on server \texttt{B}, to make server \texttt{B} have a hotspot shard and overloaded.
The throughput drops by 33\% due to load imbalancing.
At this time, server \texttt{A} and server \texttt{B} have a CPU utilization of 60.7\% and 91\% respectively.
The average CPU utilization of the other 4 servers drops to 72.8\%, since requests to overloaded server \texttt{B} suffer from long queueing and thus the limited number of clients cannot generate enough requests to other servers.
CM detects the overload after 660ms (i.e., \texttt{detect-overload}) and produces a migration task that migrates the hotspot shard from server \texttt{B} to server \texttt{A}.
The migration takes 1346ms and moves about 1.1 million objects.
The throughput increases as the migration proceeds, since more \kvget requests to the hotspot shard can be served by server ~\texttt{A}.
Finally, the system achieves a load-balanced state with steady throughput.

\subsection{Comparison with Other Systems}
\label{eval:existingKV}

We compare \rowankv with two state-of-the-art replicated KVSs designed for RDMA networks:
\begin{itemize}[ itemsep=0pt, parsep=0pt, labelsep=5pt, 
	leftmargin=*, topsep=0pt,partopsep=0pt]
    \item \textbf{Clover}~\cite{ATC20Clover}.
      Clover runs on disaggregated PM, where PM servers do not have compute resources.
    Clients perform \kvget operations via RDMA \rread verbs, and perform \kvput operations (including replication) using a combination of RDMA \rwrite and \ratomic.
    \item  \textbf{HermesKV}~\cite{ASPLOS20Hermes}. It is a DRAM-resident KVS built on 
     Hermes~\cite{ASPLOS20Hermes}, a broadcast-based replication protocol.
     HermesKV uses RPC for all inter-server communication (including replication).
     We modify the code to support PM: 
     we store objects in PM and issue \texttt{ntstore} instructions for durability; indexes are in DRAM.
     In addition, we let clients generate KV requests to HermesKV servers.

\end{itemize}

\noindent
We use ZippyDB objects and 4KB objects to test KVSs under small writes and large writes, respectively.
The key distribution follows Zipfian with parameter 0.99.
The replication factor is 3 and HermesKV runs with enabled DDIO.

\begin{figure}[t!]

  \centering
  \includegraphics[width=0.8\linewidth]{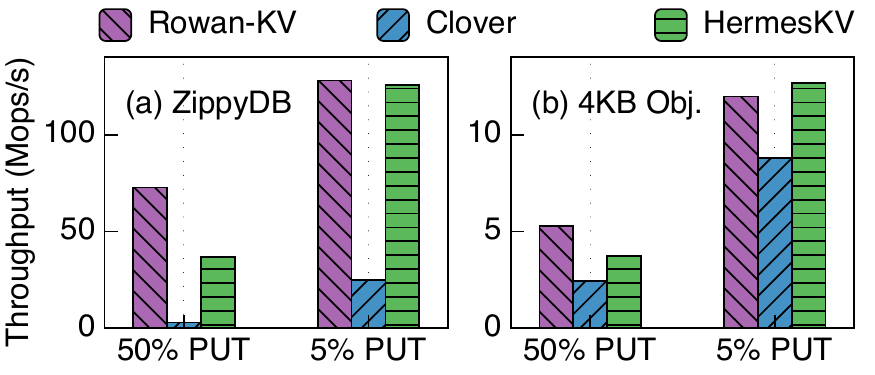}
  
  \vspace{-0.3cm}
  \caption{Comparison with Clover and HermesKV. \emph{\textbf{(a)} Throughput with ZippyDB objects. \textbf{(b)} Throughput with 4KB objects. }}
  \vspace{-0.3cm}
  \label{fig:exsitingKV}
\end{figure}

Figure~\ref{fig:exsitingKV}(a) shows the results of small writes (ZippyDB objects).
Under write-intensive workloads (i.e., 50\% \kvput),
\rowankv outperforms Clover and HermesKV by 24.5$\times$ and 1.98$\times$, respectively.
Two reasons contribute to Clover's low throughput.
First, due to the disaggregated architecture, every operation in Clover needs multiple network communications.
Second, Clover uses RDMA \ratomic to resolve conflicts between client threads, which leads to significant performance degradation when contention appears~\cite{grantnetwork}.
Using RDMA \ratomic on PM is also considered slow due to its read-modify-write behavior~\cite{ATC21PMRDMA}.
HermesKV uses RPC for replication which consumes CPU cycles at backups, so it is outperformed by \rowan which uses one-sided \rowan for replication.
We measure \DLWA of these KVSs. 
Clover has 1.86$\times$ \DLWA and HermesKV has 2.95$\times$ \DLWA, since both of them generate a large number of random small writes on PM:
for a \kvput operation, Clover performs copy-on-write using \rwrite and HermesKV performs in-place updates.
In contrast, \rowankv adopts the log-structured approach to manage objects and exploits \rowan abstraction to minimize the number of write streams;
thus, \DLWA of \rowankv is less than 1.032$\times$.
Under read-intensive workloads  (i.e., 5\% \kvput), \rowankv and HermesKV have similar throughput,
which far exceeds that of Clover (about 5$\times$).

Figure~\ref{fig:exsitingKV}(b) reports the results of large writes (4KB objects).
Under write-intensive workloads, \rowankv outperforms HermesKV by 1.42$\times$ and is bottlenecked by PM write bandwidth.
HermesKV can not approach the limitation of PM write bandwidth, since its backups waste lots of CPU cycles to copy/persist large objects from RPC buffers to PM.
Under read-intensive workloads, \rowankv and HermesKV are bottlenecked by the network bandwidth (11GB/s per server), which is much lower than PM read bandwidth (18GB/s).

\section{Discussion}
\label{sec:discussion}

Although Intel killed Optane memory business for commercial reasons in summer 2022,
we believe that \rowan is still applicable to future byte-addressable storage devices.
For example, CXL storage devices (e.g., Samsung's Memory-Semantic SSD~\cite{Samsung}), which are considered promising alternatives to Optane DIMMs, share similarities with Optane DIMMs:
1) limited write bandwidth; 2) byte interfaces with a block-level internal access granularity (e.g., flash page).
Thus, when many remote threads concurrently access CXL storage devices with small IO size, \rowan can still effectively mitigate \DLWA and thus boost system performance.
\section{Related Work}
\label{sec:related}

\smalltitle{PM KVSs}
There are a host of works on PM KVSs,
but most of them are single-machine (except Clover~\cite{ATC20Clover}) .
HiKV~\cite{ATC17HiKV} and Bullet~\cite{ATC18Bullet} are designed before the availability of real PM devices; both of them store objects into fine-grained PM hash tables.
However, real PM devices have block-level internal access granularity (e.g., 256B in Optane DIMMs).
To reduce \DLWA, recent PM KVSs, including FlatStore~\cite{ASPLOS20FlatStore}, Viper~\cite{VLDB21Viper}, and Pacman~\cite{ATC22Pacman}, adopt log-structured approaches to manage objects. 
\rowankv also uses log-structured approach for the same reason, but focuses on distributed environments where objects are sharded and replicated.

\smalltitle{RDMA replication}
RDMA replication can be categorized into two
groups, namely \emph{backup-active} and \emph{backup-passive},
depending on whether backups consume CPUs on the critical path of replication.
Lots of systems~\cite{ASPLOS15Mojim, SoCC17APUS, OSDI20Assise, ASPLOS20Hermes, DSN18RDMC, TOCS19Derecho} belong to backup-active group, where backups' CPUs need to process messages during replication.
For backup-passive group~\cite{SOSP15FaRMTX,HPDC15DARE, ATC18Tailwind,SIGCOMM18Hyperloop, CoNEXT19Sift, OSDI18DrTMH, OSDI20Mu}, 
primaries only need to wait for ACKs from the RNIC hardware of backups.
For example, Hyperloop~\cite{SIGCOMM18Hyperloop} uses RDMA \texttt{WAIT} and \rwrite verbs to realize chain replication.
\rowankv belongs to the backup-passive group, so it features low latency and high CPU efficiency.
Yet, traditional backup-passive approaches can lead to \DLWA on PM KVSs,
driving us to design the \rowan abstraction.

\smalltitle{RDMA abstraction}
Due to limited expressivity of RDMA verbs, several works propose new RDMA abstractions~\cite{EuroSys20StRoM,NSDI20FileMR,HotNets20RMC, SOSP21PRISM,MICRO49SABRes, HotOS19FarMemory}.
StRoM~\cite{EuroSys20StRoM} and RMC~\cite{HotNets20RMC} allow applications to define functions on NICs.
Aguilera et al.~\cite{HotOS19FarMemory} and PRISM~\cite{SOSP21PRISM} propose several new RDMA verbs to support far memory data structures and distributed systems.
RedN~\cite{NSDI22RedN} makes RDMA Turing complete using self-modifying chains.
All above works (except RedN) require RNIC modification or specialized hardware (e.g., SmartNICs).
In contrast, \rowan can be realized with off-the-shelf RNICs, 
leveraging RNIC features such as SRQ and MP RQ.
Besides, \rowan targets handling high fan-in small PM writes.
\section{Conclusion}

This paper explored how to efficiently replicate PM KVSs using RDMA.
We showed that existing approaches using RDMA \rwrite cause severe device-level write amplification (\DLWA) on PM.
To this end, we proposed \rowan, a one-sided RDMA abstraction that can merge numerous remote writes into a single stream.
Based on \rowan, we built \rowankv, a log-structured PM KVS;
it outperforms RPC and RDMA \rwrite alternatives in throughput and latency under write-intensive workloads, while achieving low \DLWA.

\bibliographystyle{plain}
\bibliography{paper}

\end{document}